%% file: article.tex
\documentclass[preprint,5p,times]{elsarticle}

\usepackage{amssymb}

\usepackage{booktabs}
\usepackage[table,xcdraw]{xcolor}
\usepackage{adjustbox}

\usepackage{comment}
\usepackage[justification=centering]{caption} 
\usepackage{amsmath}
\usepackage{graphicx,lmodern}
\usepackage{booktabs}
\usepackage{lipsum}
\usepackage{adjustbox}
\usepackage[ruled,vlined,onelanguage]{algorithm2e}
\usepackage{tkz-graph}
\usepackage{array}
\usepackage{blkarray}
\usepackage[justification=centering]{caption}
\usepackage{float}
\DeclareMathOperator*{\argmax}{arg\,max}
\DeclareMathOperator*{\argmin}{arg\,min}
\usepackage{url,multirow,morefloats,floatflt,cancel,tfrupee}
\usepackage{caption}
\usepackage{subcaption}

\journal{Internet of Things}

\begin{document}

\begin{frontmatter}



\title{Internet of Smart-Cameras for Traffic Lights Optimization in Smart Cities}


\author[1]{Willy Carlos Tchuitcheu
}\ead{twilly@aims.ac.rw}
\author[2]{Christophe Bobda}\ead{cbobda@ece.ufl.edu}
\author[2]{Md Jubaer Hossain Pantho
}\ead{mpantho@ufl.edu}


\cortext[cor1]{Corresponding author}
\address[1]{Camertronix, Yaounde, Cameroon}
\address[2]{Department of Electrical and Computer Engineering,University of Florida, USA}

\begin{abstract}
Smart and decentralized control systems have recently been proposed to handle the growing traffic congestion in urban cities. Proposed smart traffic light solutions based on Wireless Sensor Network and Vehicular Ad-hoc NETwork are either unreliable and inflexible or complex and costly. Furthermore, the handling of special vehicles such as emergency is still not viable, especially during busy hours. Inspired by the emergence of distributed smart cameras, we present a novel approach to traffic control at intersections. Our approach uses smart cameras at intersections along with image understanding for real-time traffic monitoring and assessment. Besides understanding the traffic flow, the cameras can detect and track special vehicles and help prioritize emergency cases. Traffic violations can be identified as well and traffic statistics collected. In this paper, we introduce a flexible, adaptive and distributed control algorithm that uses the information provided by distributed smart cameras to efficiently control traffic signals. Experimental results show that our collision-free approach outperforms the state-of-the-art of the average user's waiting time in the queue and improves the routing of emergency vehicles in a cross congestion area.

\end{abstract}




\end{frontmatter}


\input{section1_Intro.tex}

\input{section2-DesignAndModeling.tex}

\input{section3-ImageProcessing.tex}

\input{section4-WaitingQueue.tex}

\input{section5-DistributedAlgo.tex}

\input{section5_Evaluation.tex}

\input{section6-Conclusion.tex}



  \bibliographystyle{elsarticle-num} 
  \bibliography{reference}





\end{document}

%% file: section1_Intro.tex
\section{Introduction}
\label{sec:introduction}
\begin{figure}[!h]
\centering
 \captionsetup{justification=centering}
\includegraphics[width=0.8\linewidth]{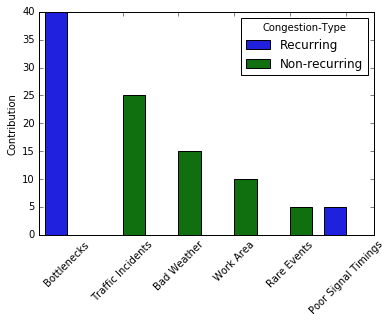}
\caption{Traffic congestion contribution as a percentage of congestion source.}
\label{sourceCongestionByType}
\end{figure}

The rapid growth in urbanization is leading to a tremendous increase in automobiles in cities\cite{hopkins2019investigating}. Unfortunately, infrastructure development has not kept up with the growth in transportation. With lack and limited availability of public transportation, traffic congestion on public roads during rush hours has become a critical problem in many countries. This problem will be unmanageable if no effort is undertaken \cite{Wan2017}\cite{Pattanaik2016}. Congestion results from traffic demand that approaches or exceeds the capacity of the available infrastructure. There are essentially two categories of traffic congestion: 1) \textbf{recurring} traffic congestion that appears at the same place and the same time every day and 2) \textbf{non-recurring} traffic congestion caused by a random unplanned event or temporary disruptions that take away part of the roadway.
The US Federal Highway Administration defines six sources of congestion\cite{fhwa20132013} as shown in Table \ref{SourceCongestion}. Figure \ref{sourceCongestionByType} shows the sources of congestion and their contribution to congestion in percent on y-axes. 

\begin{table}[!h]
\scriptsize
\caption{Congestion sources and terms.}
\label{SourceCongestion}
\centering
\begin{tabular}{@{}ll@{}}
\toprule
\textbf{Termed} & \textbf{Source } \\ \midrule
Bottlenecks & \begin{tabular}[c]{@{}l@{}}Road with inadequate physical capacity   \\ (roadway narrows, enclave).\end{tabular} \\ \\
Traffic incidents & Vehicles crashes and stalls. \\ \\
Work area & \begin{tabular}[c]{@{}l@{}} Road repairs, building of new roads and \\ maintenance activities. \end{tabular} \\ \\
Bad weather & Flood, snowfall and fog. \\ \\
Rare Events & Strikes and marathons. \\ \\
Poor signal timings & \begin{tabular}[c]{@{}l@{}}Empty lane with green light \cite{srivastava2013intelligent} and time allocated \\  with respect to the volume of the traffic on the  lane.\end{tabular} \\ \bottomrule
\end{tabular}
\end{table}
Recurring and non-recurring traffic congestion 
contribute to urban traffic congestion at almost the same rate. Heavy traffic congestion leads to waste of time, increase pollution, waste of fuel, increased cost of transportation and driving-related stress, inefficient supply chains \cite{hopkins2014alleviating}, with an adverse effect on the economy \cite{fhwa20132013}\cite{emo2016slow}\cite{loong2017time}.
Intelligent Traffic Management System can alleviate traffic congestion by 1) collecting traffic data in real-time at intersection, for instance through the use of Wireless Sensor Networks (WSNs) \cite{Sherif2014}, RFIDs, ZigBee \cite{zigbee2010}, Vehicular ad-hoc NETwork(VANETs) \cite{VANET2019}, Bluetooth devices and cameras and infrared signals, whereas WSNs have gained increasing attention in traffic detection and avoiding road congestion \cite{nellore2016survey}; 2) using adaptive algorithm to control traffic with the goal of minimizing the average waiting time in queues of users; 3) incorporating a mechanism to allow emergency vehicles to easily cross congestion areas.

Despite various techniques and research for alleviating traffic congestion including "anywhere working" \cite{hopkins2019investigating}, "Markov chain traffic assignment" \cite{salman2018alleviating}, and government's policies such as congestion pricing, driving restrictions, vehicle purchase restrictions, and public transit investment, the Intelligent Traffic Management System (ITMS) \cite{ITMS2014} continues to face  significant challenges  namely: 
\begin{itemize}
\item  Congestion : can ITMS  react quickly to non-recurring congestion problem.
\item  Traffic incident notification (Traffic violations, traffic rules) : can  ITMS  send real-time information to police  to act upon the situation.
\item How to maintain coordination between intersections for a safety smart city.
\end{itemize}

Table \ref{summaryWSN} presents several technologies used for traffic control. Most of those systems rely  on Wireless Sensor Network (WSN) for traffic control coupled with cameras for video surveillance \cite{Franceschinis2009}. These methods are tedious and involve a large amount of hardware. 

\begin{figure}[!h]
\centering
 \captionsetup{justification=centering}
\includegraphics[width=0.8\linewidth]{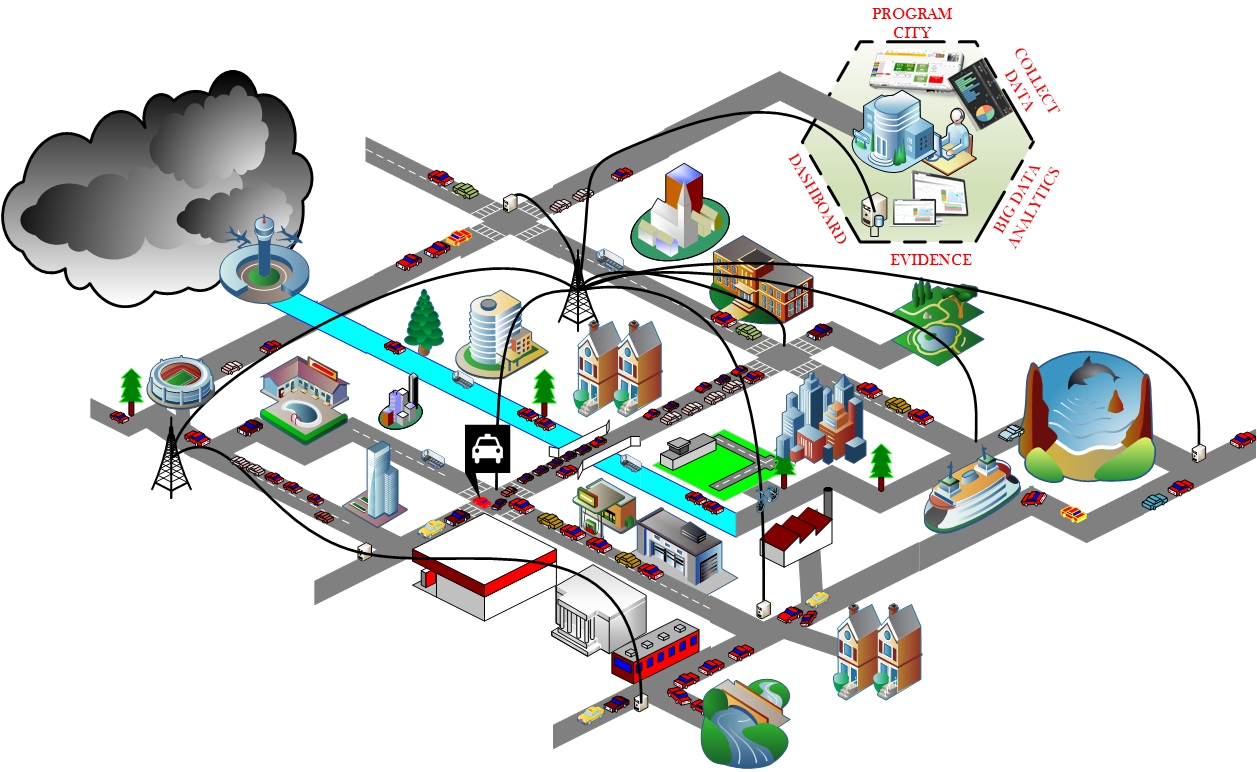}
\caption{Smart city configuration with smart infrastructure, cloud-based data collection and decision making.}
\label{smartCity}
\end{figure}

In this paper, we propose a new and versatile method that uses distributed smart cameras along with advance image understanding to supply the waiting queue in real-time with traffic data (vehicles count, types, density, etc...). This information can then be used by a central authority to control the whole traffic infrastructure. Figure \ref{smartCity} shows an example of smart city with multiple connected infrastructure, including traffic light.

Data collected from those infrastructures are sent to the could for analysis and decision making. In the specific case of traffic, data collected from all traffic light can be used for inter-intersection control, tracking of special vehicles, traffic offenders and address the non-recurring congestion area by changing traffic rules of the adjacent intersections. The current paper focuses on smart control at road-intersections. The global management of the infrastructure and decision making in the whole context of a smart city is not in the scope of this work and will be addressed in future research.

The rest of the paper is organized as follows. Section \ref{sec:related} discusses the related work. In section \ref{sec:design}, we describe our modeling system. In sections \ref{sec:algorithm}, \ref{sec:wait}, and \ref{sec:distributed}, we explain our implemented algorithm, and the simulation and results in section \ref{sec:simulation}.

\section{ Related Works}\label{sec:related}
In this section, we discuss related work on smart traffic control. We divide the work in two categories: data collection and traffic control solutions.

\subsection{Traffic Data Collection Technologies}
In \cite{padmavathi2010study}, a review of technologies for traffic data at the intersections is performed. For each technology, the authors consider the ability to count vehicles, detect vehicles, detect the speed of vehicles and the ability to distinguish different vehicle types. The authors also consider whether the technology could be used in a  multi-lane scenario, and the bandwidth required to communicate the information to a central control. The summary of the survey is shown in table  \ref{summaryWSN}.

\begin{table*}[!h]
\caption{Technology along with traffic data collected \cite{padmavathi2010study}.}
\centering
\scriptsize
\begin{tabular}{@{}llllllll@{}}
\toprule
\multicolumn{1}{c}{\textbf{Technology}} & \multicolumn{1}{c}{\textbf{\begin{tabular}[c]{@{}c@{}}Vehicle \\ Count\end{tabular}}} & \multicolumn{1}{c}{\textbf{Presence}} & \multicolumn{1}{c}{\textbf{Speed}} & \multicolumn{1}{c}{\textbf{\begin{tabular}[c]{@{}c@{}}Output \\ Data\end{tabular}}} & \multicolumn{1}{c}{\textbf{Classification}} & \multicolumn{1}{c}{\textbf{\begin{tabular}[c]{@{}c@{}}Multiple Lane, \\ Multiple Detection\\  Zone Data\end{tabular}}} & \multicolumn{1}{c}{\textbf{\begin{tabular}[c]{@{}c@{}}Communication\\  Bandwidth\end{tabular}}} \\ \midrule
\textbf{Inductive loop} & $\surd$ & $\surd$ & $\surd$ * & $\surd$ & $\surd$ \& &  & Low to modest \\
\textbf{Magnetometer} & $\surd$ & $\surd$ & $\surd$ * & $\surd$ &  &  & Low \\
\textbf{Magnetic induction coil} & $\surd$ & $\surd$ \$ & $\surd$ * & $\surd$ &  &  & Low \\
\textbf{Microwave radar} & $\surd$ & $\surd$  \# & $\surd$ & $\surd$ \# & $\surd$ \# & $\surd$ \# & Moderate \\
\textbf{Active infrared} & $\surd$ & $\surd$ & $\surd$ @ & $\surd$ & $\surd$ & $\surd$ & Low to modest \\
\textbf{Passive infrared} & $\surd$ & $\surd$ & $\surd$ @ & $\surd$ &  &  & Low to modest \\
\textbf{Ultrasonic} & $\surd$ & $\surd$ &  & $\surd$ &  &  & Low \\
\textbf{Acoustic array} & $\surd$ & $\surd$ & $\surd$ & $\surd$ &  & $\surd$  \textasciicircum{} & Low to modest \\
\textbf{Video image processor} & $\surd$ & $\surd$ & $\surd$ & $\surd$ & $\surd$ & $\surd$ & Low to high \\
 &  &  &  &  &  &  &  \\
\multicolumn{8}{c}{\begin{tabular}[c]{@{}c@{}}* Two sensors can be used to measure speed; \& With specific electronics device that categorizes vehicles;\\  \$ By using distinct sensor layouts and data processing software; \# By using a microwave radar sensor and suitable signal processing unit; \\ @ With multi detection region; \textasciicircum By suitable beam forming models and data processing unit\end{tabular}.} \\ \bottomrule
\end{tabular}
\label{summaryWSN}
\end{table*}

 The presented technologies are not suitable to detect special vehicles such as emergency and police cars. Furthermore, to detect the entire intersection, multiple sensor nodes must be deployed to increase coverage. As a result, efficient coordination with the central system becomes very challenging. Our approach in this work is base on video, which is free of the previous-mentioned challenges.

\subsection{Traffic Control Algorithm }
There is a large amount of literature on the subject of traffic control algorithms. We provide an overview of the representative examples. \par In \cite{faye2012distributed}, the authors propose WSN architecture and an algorithm for controlling green lights on a single intersection. Their solution is designed for an isolated intersection and reduces waiting times without introducing congestion.  
 \par  The approach has been extended in \cite{faye2012distributed2} to an algorithm called TAPIOCA
\textit{(distribuTed and Adaptative IntersectiOns Control Algorithm)} that considers multiple adjacent intersections that communicate with adjacent intersections using WSN. \par In \cite{faye2014multiple}, an improvement of TAPIOCA was proposed for the multi-intersection case by defining mechanisms to ease offloading between close intersections and to create green waves. In \cite{wu2015distributed}, a novel approach to traffic control at the intersection was proposed. Rather than solving the optimization problem of green light scheduling, vehicles compete for the privilege of passing by exchanging messages. In this case, the vehicles must have an extra device that allows them to communicate. \par In \cite{wongpiromsarn2012distributed}, an adaptive algorithm was proposed from the back-pressure routing, which has been mainly applied to communication and power networks. In \cite{yousef2010intelligent}, an adaptive traffic control for both single and multiples intersections was proposed. \par In \cite{nellore2016survey} a survey has been made on adaptive algorithms for traffic control. However, these approaches haven't considered the priority for emergency vehicles, and in case of changing traffic rules or maintenance at the intersection, the algorithm fails to adapt. 

In general, algorithms that rely on WSN for collection of traffic data are not resilient to certain critical scenarios such as changing the highway code at the intersection, maintenance work on pavement. Furthermore, they do not provide a safety gateway to act under certain critical cases occurring at the intersection remotely. The approach we present in this work is resilient capable of withstanding several critical cases that the existing approach cannot handle.

%% file: section2-DesignAndModeling.tex
\section{Design and Modeling}
\label{sec:design}
In this section, we will first describe the component of traffic light system we intend to optimize. A description of our overall control architecture will follow.

\subsection{Ecosystem}
The intersections considered in this way are 4-way right-side driving intersections. Each way has the three-color traffic light located at the right top and a smart camera installed in the face of the road. All possible movements are allowed.  An intersection is where multiple roads cross. A road is a set of lanes. 
\begin{figure*}[!h]

\centering 
 \captionsetup{justification=centering}
\includegraphics[width=1\linewidth]{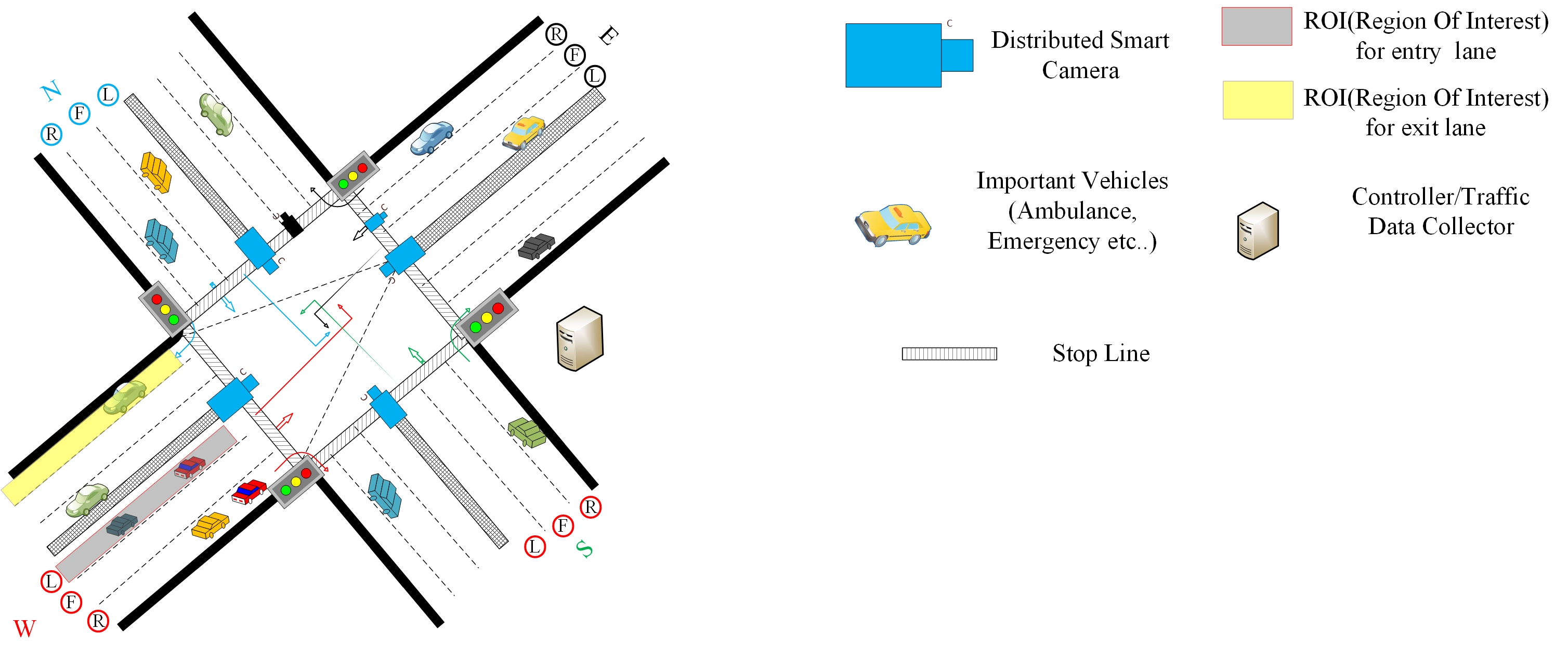}
\caption{Distributed Smart-Camera based ITMS.}
\label{Eco} 
\end{figure*}  
Figure \ref{Eco} shows an intersection with four lanes marked N (North), S(South), W (West) and E (East) that intersect. Each path has three lanes in the incoming direction, which are turn-left(L), go-forward (F) and turn-right(R). A passing vehicle can have a path P of $\{E,S,N,W\}$ and a direction D of $\{L,F,R\}$. Thus, a lane that has a vehicle can be determined by a pair of $\{P,D\}$. There are twelve lanes with labels pair $(P,D):\{WR, WF, WL, ER, EF, EL, NR, NF, NL, SR, SF, SL\}$.  
Modeling the intersection to comply with road regulations is provided in the following table. 
\begin{table}[!h]
\scriptsize
\caption{Matrix of conflict.}
\label{table:conflictMatrix}
\centering
\begin{adjustbox}{width=.5\textwidth}
\begin{tabular}{ccccccccccccc}
\hline
\multicolumn{1}{|c|}{\textbf{}} & \multicolumn{1}{c|}{\textbf{WR}} & \multicolumn{1}{c|}{\textbf{WF}} & \multicolumn{1}{c|}{\textbf{WL}} & \multicolumn{1}{c|}{\textbf{ER}} & \multicolumn{1}{c|}{\textbf{EF}} & \multicolumn{1}{c|}{\textbf{EL}} & \multicolumn{1}{c|}{\textbf{NR}} & \multicolumn{1}{c|}{\textbf{NF}} & \multicolumn{1}{c|}{\textbf{NL}} & \multicolumn{1}{c|}{\textbf{SR}} & \multicolumn{1}{c|}{\textbf{SF}} & \multicolumn{1}{c|}{\textbf{SL}} \\ \hline
\multicolumn{1}{|c|}{\textbf{WR}} & \multicolumn{1}{c|}{\cellcolor[HTML]{FE0000}} & \multicolumn{1}{c|}{} & \multicolumn{1}{c|}{} & \multicolumn{1}{c|}{} & \multicolumn{1}{c|}{} & \multicolumn{1}{c|}{$\otimes$} & \multicolumn{1}{c|}{} & \multicolumn{1}{c|}{$\otimes$} & \multicolumn{1}{c|}{} & \multicolumn{1}{c|}{} & \multicolumn{1}{c|}{} & \multicolumn{1}{c|}{} \\ \hline
\multicolumn{1}{|c|}{\textbf{WF}} & \multicolumn{1}{c|}{} & \multicolumn{1}{c|}{\cellcolor[HTML]{FE0000}} & \multicolumn{1}{c|}{} & \multicolumn{1}{c|}{} & \multicolumn{1}{c|}{} & \multicolumn{1}{c|}{$\otimes$} & \multicolumn{1}{c|}{} & \multicolumn{1}{c|}{$\otimes$} & \multicolumn{1}{c|}{$\otimes$} & \multicolumn{1}{c|}{$\otimes$} & \multicolumn{1}{c|}{$\otimes$} & \multicolumn{1}{c|}{$\otimes$} \\ \hline
\multicolumn{1}{|c|}{\textbf{WL}} & \multicolumn{1}{c|}{} & \multicolumn{1}{c|}{} & \multicolumn{1}{c|}{\cellcolor[HTML]{FE0000}} & \multicolumn{1}{c|}{$\otimes$} & \multicolumn{1}{c|}{$\otimes$} & \multicolumn{1}{c|}{} & \multicolumn{1}{c|}{} & \multicolumn{1}{c|}{$\otimes$} & \multicolumn{1}{c|}{$\otimes$} & \multicolumn{1}{c|}{} & \multicolumn{1}{c|}{$\otimes$} & \multicolumn{1}{c|}{$\otimes$} \\ \hline
\multicolumn{1}{|c|}{\textbf{ER}} & \multicolumn{1}{c|}{} & \multicolumn{1}{c|}{} & \multicolumn{1}{c|}{$\otimes$} & \multicolumn{1}{c|}{\cellcolor[HTML]{FE0000}} & \multicolumn{1}{c|}{} & \multicolumn{1}{c|}{} & \multicolumn{1}{c|}{} & \multicolumn{1}{c|}{} & \multicolumn{1}{c|}{} & \multicolumn{1}{c|}{} & \multicolumn{1}{c|}{$\otimes$} & \multicolumn{1}{c|}{} \\ \hline
\multicolumn{1}{|c|}{\textbf{EF}} & \multicolumn{1}{c|}{} & \multicolumn{1}{c|}{} & \multicolumn{1}{c|}{$\otimes$} & \multicolumn{1}{c|}{} & \multicolumn{1}{c|}{\cellcolor[HTML]{FE0000}} & \multicolumn{1}{c|}{} & \multicolumn{1}{c|}{$\otimes$} & \multicolumn{1}{c|}{$\otimes$} & \multicolumn{1}{c|}{$\otimes$} & \multicolumn{1}{c|}{} & \multicolumn{1}{c|}{$\otimes$} & \multicolumn{1}{c|}{$\otimes$} \\ \hline
\multicolumn{1}{|c|}{\textbf{EL}} & \multicolumn{1}{c|}{$\otimes$} & \multicolumn{1}{c|}{$\otimes$} & \multicolumn{1}{c|}{} & \multicolumn{1}{c|}{} & \multicolumn{1}{c|}{} & \multicolumn{1}{c|}{\cellcolor[HTML]{FE0000}} & \multicolumn{1}{c|}{} & \multicolumn{1}{c|}{$\otimes$} & \multicolumn{1}{c|}{$\otimes$} & \multicolumn{1}{c|}{} & \multicolumn{1}{c|}{$\otimes$} & \multicolumn{1}{c|}{$\otimes$} \\ \hline
\multicolumn{1}{|c|}{\textbf{NR}} & \multicolumn{1}{c|}{} & \multicolumn{1}{c|}{} & \multicolumn{1}{c|}{} & \multicolumn{1}{c|}{} & \multicolumn{1}{c|}{$\otimes$} & \multicolumn{1}{c|}{} & \multicolumn{1}{c|}{\cellcolor[HTML]{FE0000}} & \multicolumn{1}{c|}{} & \multicolumn{1}{c|}{} & \multicolumn{1}{c|}{} & \multicolumn{1}{c|}{} & \multicolumn{1}{c|}{$\otimes$} \\ \hline
\multicolumn{1}{|c|}{\textbf{NF}} & \multicolumn{1}{c|}{$\otimes$} & \multicolumn{1}{c|}{$\otimes$} & \multicolumn{1}{c|}{$\otimes$} & \multicolumn{1}{c|}{} & \multicolumn{1}{c|}{$\otimes$} & \multicolumn{1}{c|}{$\otimes$} & \multicolumn{1}{c|}{} & \multicolumn{1}{c|}{\cellcolor[HTML]{FE0000}} & \multicolumn{1}{c|}{} & \multicolumn{1}{c|}{} & \multicolumn{1}{c|}{} & \multicolumn{1}{c|}{$\otimes$} \\ \hline
\multicolumn{1}{|c|}{\textbf{NL}} & \multicolumn{1}{c|}{} & \multicolumn{1}{c|}{$\otimes$} & \multicolumn{1}{c|}{$\otimes$} & \multicolumn{1}{c|}{} & \multicolumn{1}{c|}{$\otimes$} & \multicolumn{1}{c|}{$\otimes$} & \multicolumn{1}{c|}{} & \multicolumn{1}{c|}{} & \multicolumn{1}{c|}{\cellcolor[HTML]{FE0000}} & \multicolumn{1}{c|}{$\otimes$} & \multicolumn{1}{c|}{$\otimes$} & \multicolumn{1}{c|}{} \\ \hline
\multicolumn{1}{|c|}{\textbf{SR}} & \multicolumn{1}{c|}{} & \multicolumn{1}{c|}{$\otimes$} & \multicolumn{1}{c|}{} & \multicolumn{1}{c|}{} & \multicolumn{1}{c|}{} & \multicolumn{1}{c|}{} & \multicolumn{1}{c|}{} & \multicolumn{1}{c|}{} & \multicolumn{1}{c|}{$\otimes$} & \multicolumn{1}{c|}{\cellcolor[HTML]{FE0000}} & \multicolumn{1}{c|}{} & \multicolumn{1}{c|}{} \\ \hline
\multicolumn{1}{|c|}{\textbf{SF}} & \multicolumn{1}{c|}{} & \multicolumn{1}{c|}{$\otimes$} & \multicolumn{1}{c|}{$\otimes$} & \multicolumn{1}{c|}{$\otimes$} & \multicolumn{1}{c|}{$\otimes$} & \multicolumn{1}{c|}{$\otimes$} & \multicolumn{1}{c|}{} & \multicolumn{1}{c|}{} & \multicolumn{1}{c|}{$\otimes$} & \multicolumn{1}{c|}{} & \multicolumn{1}{c|}{\cellcolor[HTML]{FE0000}} & \multicolumn{1}{c|}{} \\ \hline
\multicolumn{1}{|c|}{\textbf{SL}} & \multicolumn{1}{c|}{} & \multicolumn{1}{c|}{$\otimes$} & \multicolumn{1}{c|}{$\otimes$} & \multicolumn{1}{c|}{} & \multicolumn{1}{c|}{$\otimes$} & \multicolumn{1}{c|}{$\otimes$} & \multicolumn{1}{c|}{$\otimes$} & \multicolumn{1}{c|}{$\otimes$} & \multicolumn{1}{c|}{} & \multicolumn{1}{c|}{} & \multicolumn{1}{c|}{} & \multicolumn{1}{c|}{\cellcolor[HTML]{FE0000}} \\ \hline
\textbf{} &  &  &  &  &  &  &  &  &  &  &  &  \\
\textbf{} &  & $\otimes$ & \multicolumn{2}{l}{\textit{\textbf{Not Allowed}}} &  &  &  & \cellcolor[HTML]{FE0000} & \multicolumn{3}{c}{\textit{\textbf{Illegal Case}}} & 
\end{tabular}
\end{adjustbox}
\label{MatrixCon}
\end{table}

Table \ref{MatrixCon}  shows  conflict direction matrix \cite{yousef2010intelligent} . An empty box $\{ blank\}$  means that both lanes can proceed without a possibility of collision. A crossed circle $\{ \otimes\}$ means that a collision may occur if both lanes are allowed to proceed. Because of this possibility of collision, lanes that have a crossed circle will not be allowed to proceed simultaneously.


\begin{figure}[!h]
        \centering
        \includegraphics[width=0.5\linewidth]{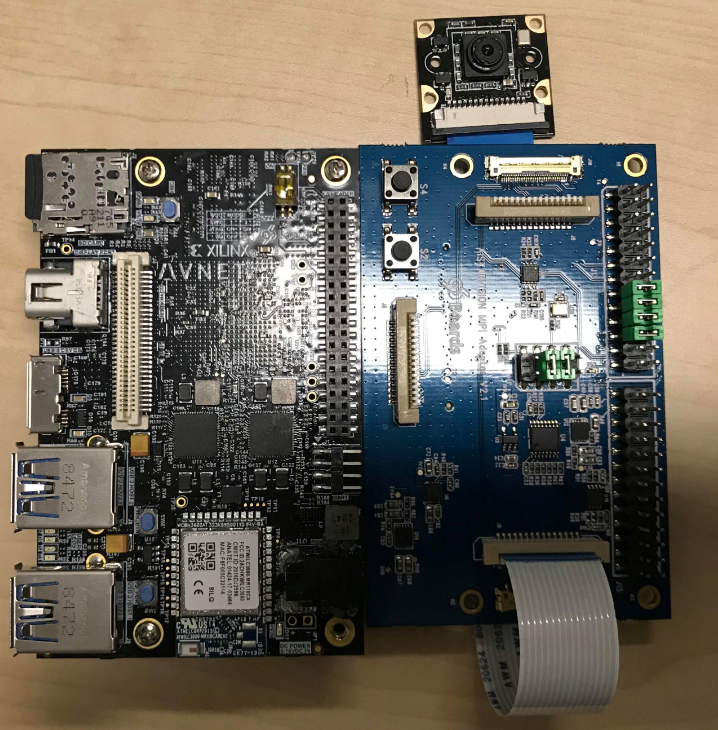}
        \caption{Smart camera setup. The ultra96 MPSoC with the camera connected to the extension board.}
        \label{fig:edge-device}
\end{figure}

\subsection{Proposed Architecture}
\begin{figure*}[!h]
        \centering
        \includegraphics[width=1\linewidth]{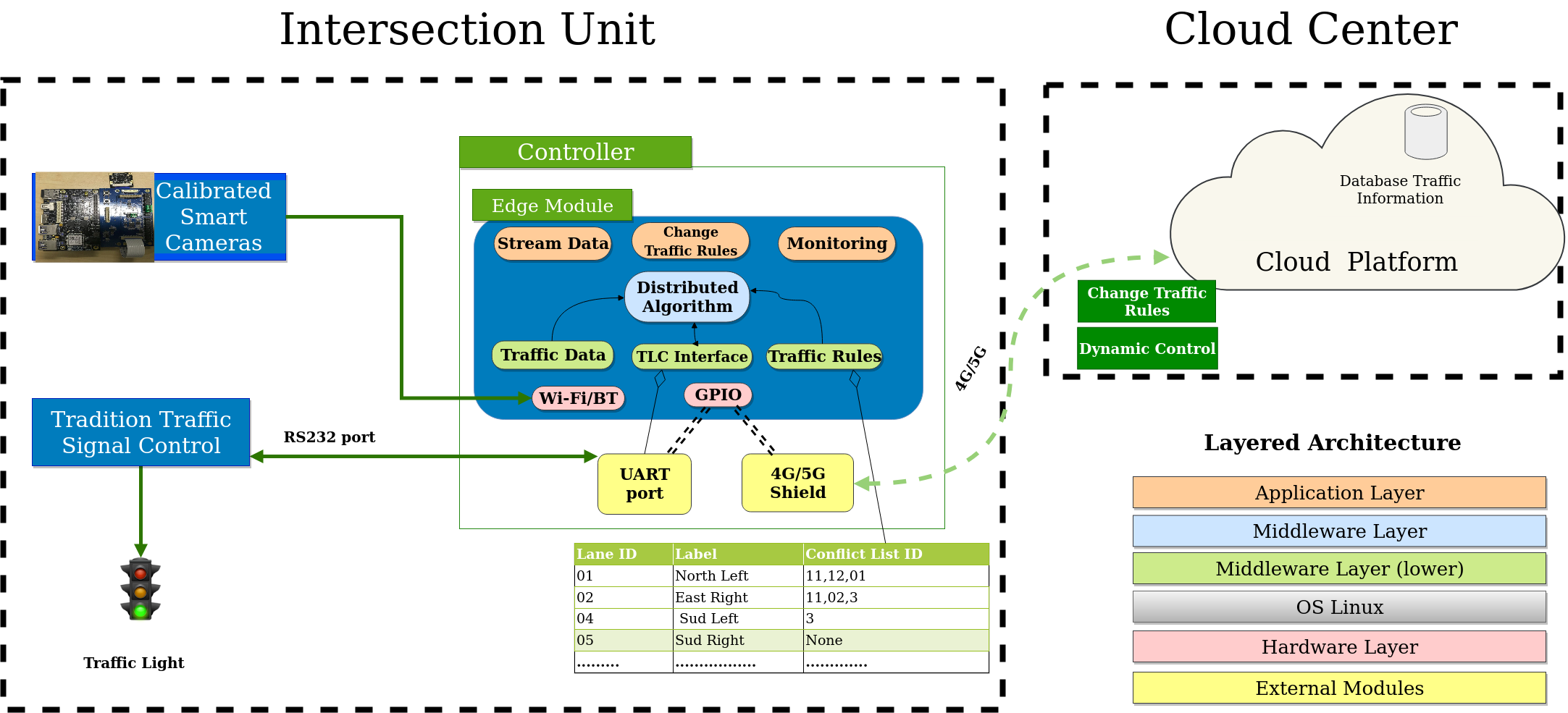}
        \caption{Proposed architecture system.}
        \label{centralizeArch}
\end{figure*}

Our proposed architecture system is shown on Figure \ref{centralizeArch}. The proposed framework comprises two components: the intersection unit and the cloud center. Within the intersection, there are three sub-components namely the controller, calibrated smart cameras, and the traditional traffic signal control unit. We designed each smart camera by connecting a MIPI CSI-2 camera to a Zynq UltraScale+ MPSoC board (ZU3EG) \cite{xilinxMPSoC}. The MPSoC board hosts a 64-bit Arm Cortex-A53 processor infused within the programmable logic. We designed the video pipeline on the programmable logic to receive image frames on the processor. Within the programmable logic, image data is transmitted through an AXI stream link. The embedded processor performs video traffic analyzer algorithm on image frame for extracting traffic-related knowledge (number and type of vehicles) within its calibrated space (waiting queues) in real-time. This information is sent to the controller unit through the communication medium (i.e. Wi-Fi). The edge module within the controller unit which is implemented on a similar 64-bit arm processor receives the information from different calibrated smart cameras and relies on traffic rules(matrix of conflict see table \ref{table:conflictMatrix}) for optimizing the traffic light through the traffic signal control.  The layered architecture of our controller unit indicating the data flow from the bottom external modules to the application layer is shown in figure \ref{centralizeArch}. It shows how the intersection can be remotely controlled by changing traffic rules from the cloud.


%% file: section3-ImageProcessing.tex
\section{Image Processing Module }\label{sec:algorithm}
In this section, we discuss the image processing used for traffic analytics. This module is embedded within the smart cameras at intersections and perform processing in-situ. The advantage of this approach is reducing data transport. This section describes algorithms within the camera needed to extract knowledge such as number, types, and direction of cars in each lane as well as the density of traffic.


\subsection{Detecting Vehicles}
The goal is to identify the lanes separated by the white lines on the road and to detect and count the number of vehicles in each lane. Upon installing cameras at the intersection, calibration is first performed to define the regions of the pictures corresponding to the different lanes. While this detection could be done automatically, without calibration, the computation overhead is too high for run-time in-situ computation.

Camera calibration is spatial and consists of determining what parts of the scene are viewed by each camera at the intersection \cite{bobda2014distributed}.

The algorithm works on the set of lines $Q =\{l_0, l_1,...,$$l_{2p-1} \}$ used to determine the $p$ Waiting Queues (WQ). Each line is represented by two random points $(x_1,y_1)$ and $(x_2,y_2)$ on that line: $(l_i) ~:~ a_ix+b_iy+c_i=0$ where $a_i=y_2-y_1$, $b_i=x_1-x_2$,  and  $c_i=-a_ix_1-b_iy_1.$ Each lane $L_p = (l_{2i},L_{2i+1})$ is bounded by two lines left $l_{2i}$ and right $L_{2i+1}$. Furthermore, for each lane $L_p$ we assign a unique color $C_p.$ Algorithm \ref{AlgoDetecVehicles} provides details on the vehicle counting in the lanes for the purpose of building the waiting queue.

\begin{algorithm}[!ht]
\scriptsize
\caption{Algorithm for Counting Vehicles in Waiting lines}
 \label{AlgoDetecVehicles}
 \caption{Algorithm Count Number Vehicle on each Waiting Queue}
\SetAlgoLined
\tcp{The input of this algorithm in a frame  or a picture }
 \KwData{Frame $F$ or image}
 \tcp{The output  is the number of vehicle on each Waiting queue }
\KwResult{Number of vehicle in each Lane }
 \tcp{For each frame $F$, use Mask-RCNN to detect objects and return the set of rectangles where vehicles have been detected }
 $ObjectDetect \longleftarrow Net.setInput(F)$\; 
 \tcp{Initialize the Number of vehicles to zero }
 $NumVehicle[0,...,p-1] \longleftarrow 0 $\; 
 \tcp{ Classify Objects with respect to lanes}
  \For{$rect(x_0,y_0,width,height) \in ObjectDetect$}{
  \tcp{find the center of the rectangle }
   $A(x_c,y_c) \longleftarrow (x_0+\frac{width}{2},y_0-\frac{height}{2})$\;
   \tcp{find the nearest line in $Q$ to $p$ by computing the distance }
   $j \longleftarrow \smash{\displaystyle\argmin_{i \in \{0,...,p-1\}}}d(A,l_i),~ \text{ where } d(A,l_i)=\frac{\|a_ix_c+b_iy_c+c_i \|}{\sqrt{a_i^2+b_i^2}}, ~~ l_i \in Q $\;
   \tcp{ localize the Lane of the vehicle $L_k$ }
    $k \longleftarrow quotion(j/2) $ \;
    assign the color $C_k$ to the vehicle\;
    \tcp{Increase the number of vehicle of the Lane $L_k$ }
    $NumVehicle[k] \longleftarrow NumVehicle[k] +1 $ \;
 }
\end{algorithm}

\subsection{Density Measurement}
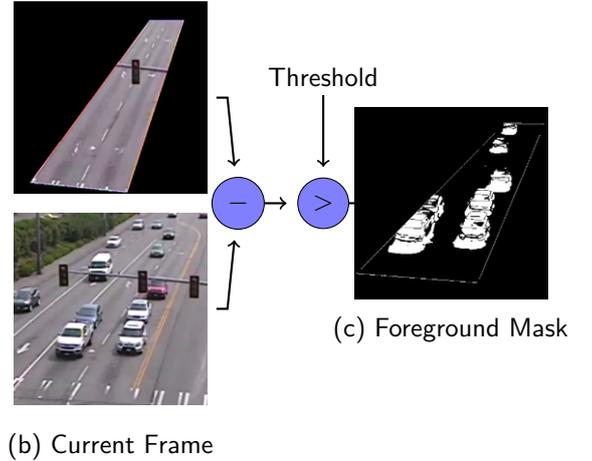
\begin{figure}[!h]
\centering
 \captionsetup{justification=centering}
\begin{tikzpicture}[scale=1.4, auto,swap, line/.style     = { draw, thick, ->, shorten >=4pt }]

 \node [text centered](labelmask) at (0,3.3) {\textsf{(a) Static Background Model}}; 
 \node (mask) at (0,2) {\includegraphics[width=1in]{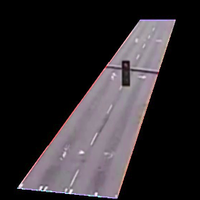}};   
\node (tTest) at (2,2.2) {\textsf{Threshold}};         
 \node [circle,draw,fill=blue!50](subtract) at (1.2,1) {$-$}; 
 \node  [circle,draw,fill=blue!50](tSign) at (2,1){$>$};
  \node [text centered](labelmask) at (3.2,-0.2) {\textsf{(c) Foreground Mask }};  
  \node  (foreground) at (3.2,1) {\includegraphics[width=1in]{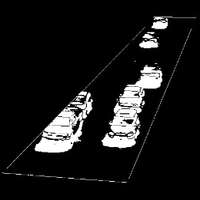}};   
   \node [text centered](labelmask) at (0,-1.3) {\textsf{(b) Current Frame}}; 
   \node  (currentFrame) at (0,0){\includegraphics[width=1in]{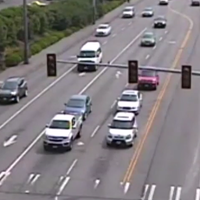}}; 
   
 \begin{scope} [every path/.style=line]
    \path (tSign)        --    (foreground);
    \path (subtract)      --    (tSign);
    \path (tTest)      --    (tSign);
     \draw[->] (mask) -- (1.1,2) -- (subtract);
     \draw[->] (currentFrame) -- (1.1,0) -- (subtract);
  \end{scope}
  
\end{tikzpicture}  
\caption{Density modeling. Figure (a) shows the static background model generated during the calibration step. (b) shows a test frame at any time t. (c) illustrates the foreground mask extracted from the image.}
\label{densityModeling}
\end{figure}

In this section, we measure in real-time the density of a queue by performing simple background subtraction.\\
\textbf{Density of a queue:} We define the density $d_a$ of a queue $\{a\}$  as the proportions of space occupied by the vehicles in $\{a\}$. 
Background Subtraction(BS) technique \cite{bradski2008learning} is used to compute the foreground mask and measure the density from the mask as shown in Figure \ref{densityModeling}
In the first step, the background model is computed when the road is free. In the second step, we are comparing the current frame to the background model in order to detect the objects(vehicles, truck, etc.)  on the scene. Since the foreground mask is a binary image $\{0=\text{Black}, 1=\text{White}\}$, we can compute the density by counting the proportion of white over that area of ROI. 
\begin{equation}
 d = \frac{countNonZero()}{Area_{ROI}}
 \label{density}
\end{equation}


Because the focus of the work is on the infrastructure for efficient control of the traffic, we will not dive into the details of the machine learning algorithms for car detection. Those algorithms are part of the available machine learning package and can be integrated into any framework.

\subsection{Implementation}
\begin{figure}[!ht]
\centering
 \captionsetup{justification=centering}
\includegraphics[width=2.5in]{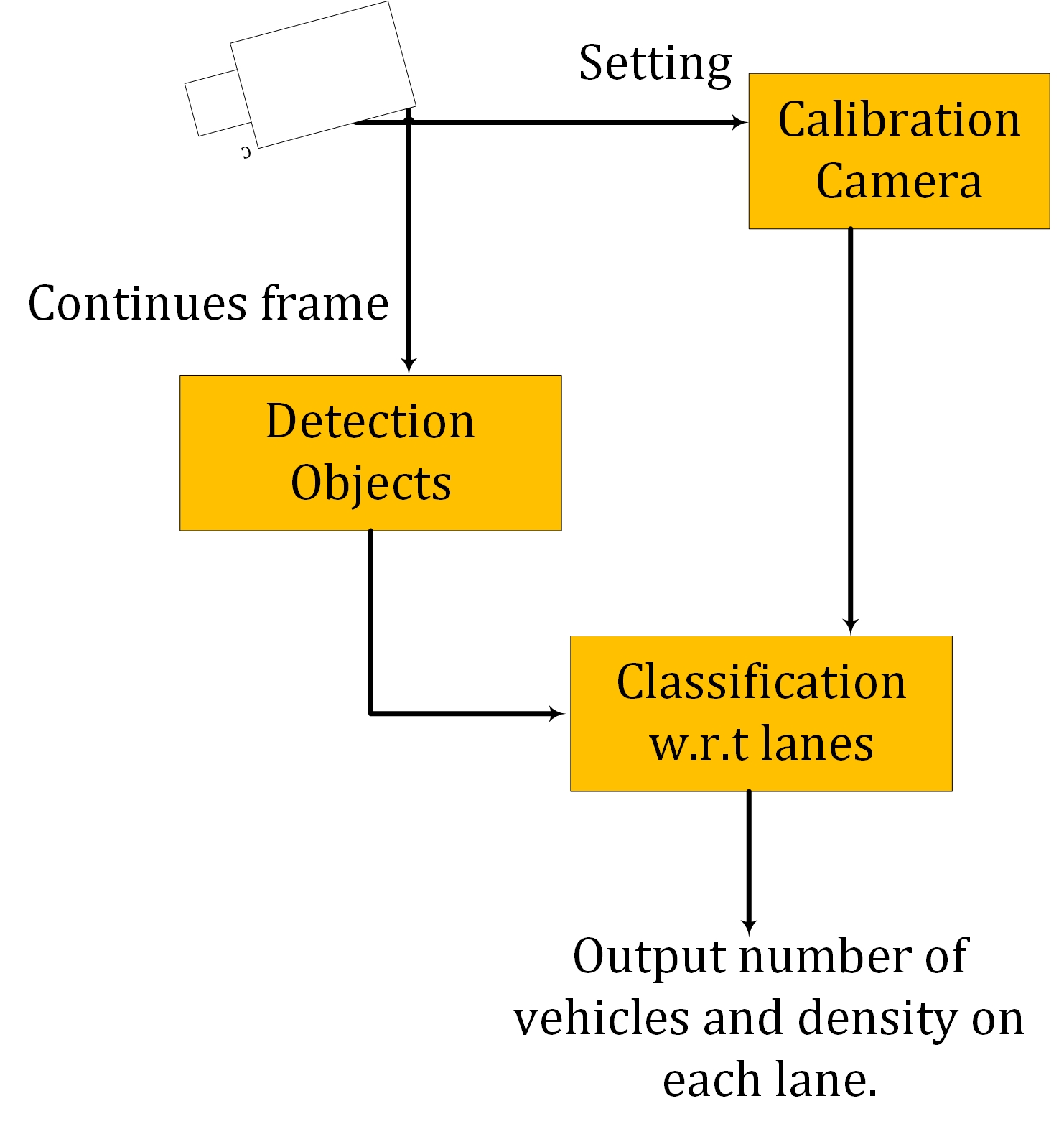}
\caption{Design implementation.}
\label{calibration}
\end{figure}

\begin{figure}[!h]
    \begin{subfigure}{.5\textwidth}
        \centering
         \captionsetup{justification=centering}
        \includegraphics[width=0.7\linewidth]{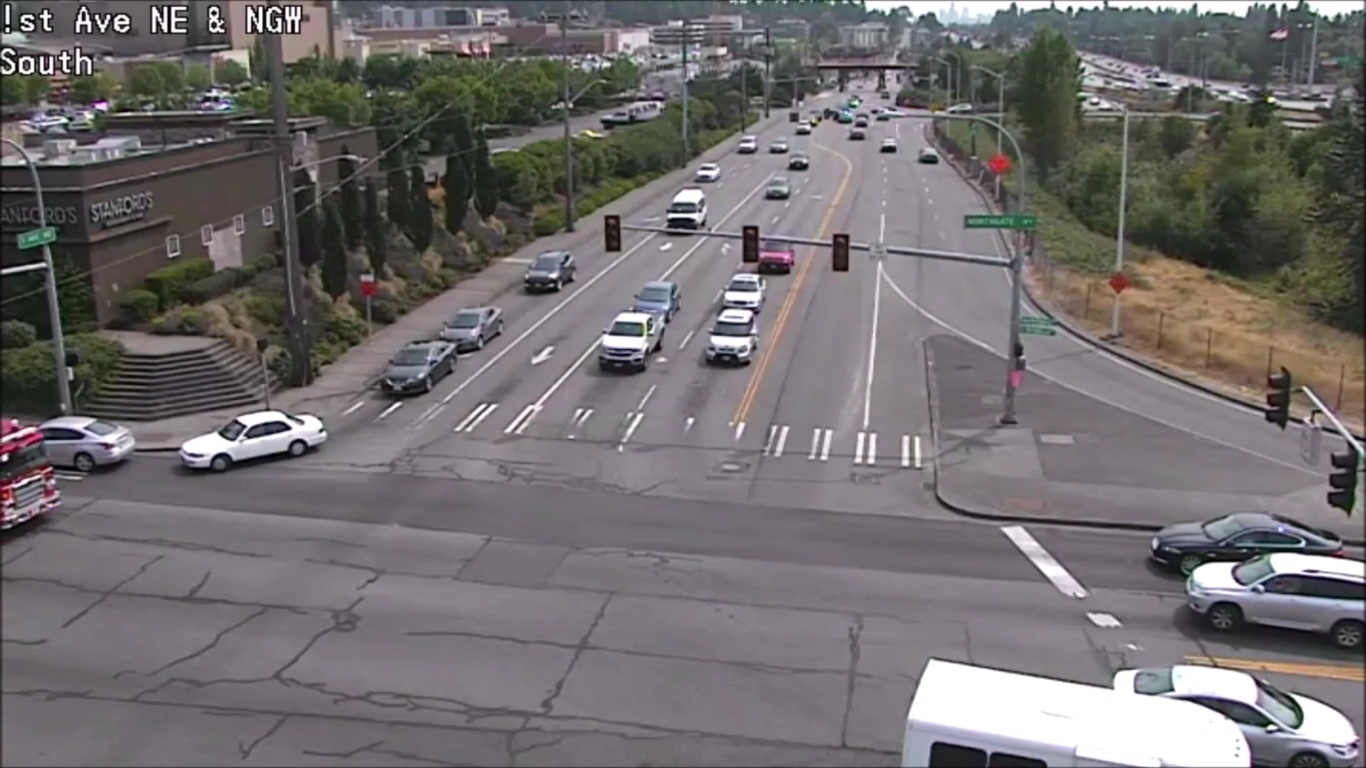}
        \caption{Frame captured before applying the algorithm.}
        \label{result1}
    \end{subfigure}
    \begin{subfigure}{.5\textwidth}
        \centering
         \captionsetup{justification=centering}
        \includegraphics[width=0.7\linewidth]{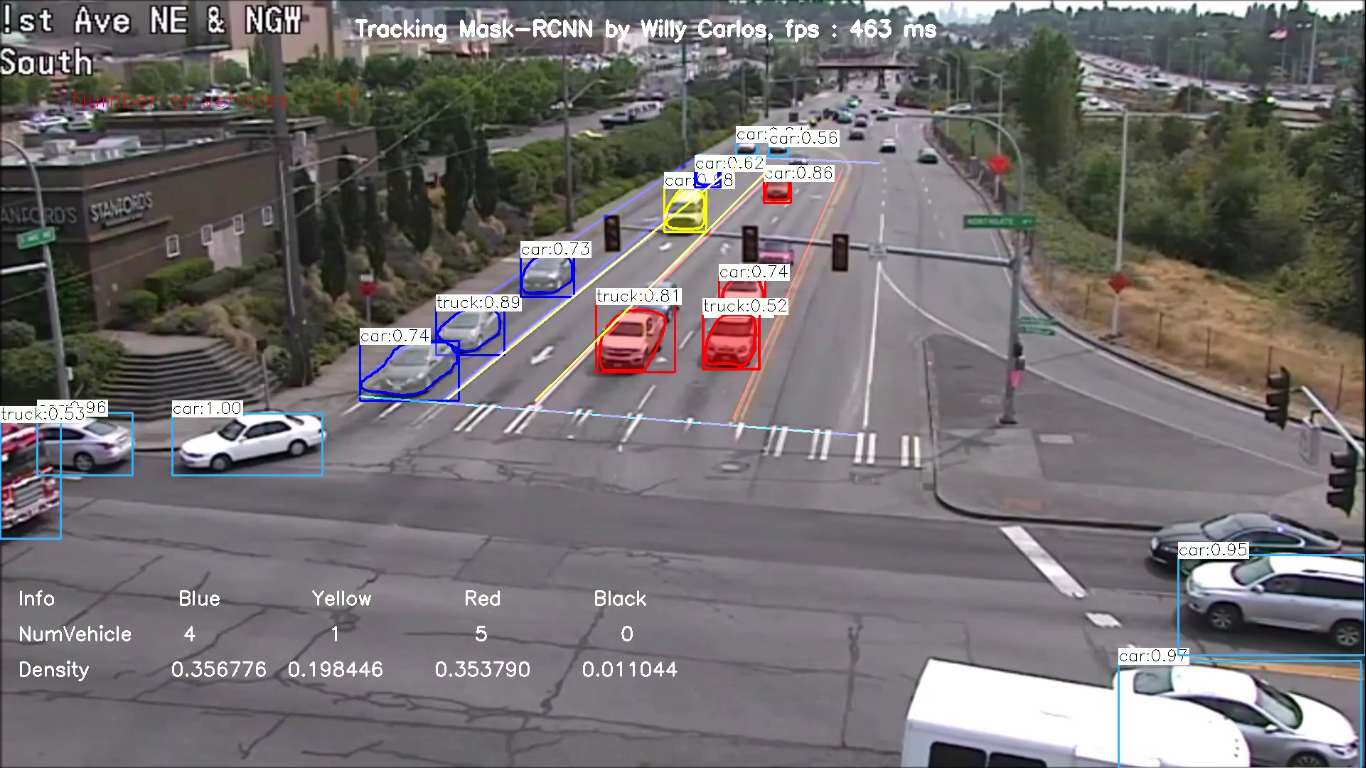}
        \caption{Output frame after applying the algorithm.}
        \label{result2}
    \end{subfigure}
     \caption{Result  produced from recorded traffic data.}
        \label{result}
\end{figure}

We implemented our algorithm based on the Region based Convolution Neural Network  \cite{MASKRCNN} to output an object label, bounding box, and the mask.  We  have used open-source OpenCV  \cite{opencv_library} version 4.0.0 in C++ language. Figure \ref{calibration} presents the flowchart of our implantation for density measurement and counting vehicles.
We calibrated the camera in space in order to set the set of foreground mask and bounded of a lane for a future  feature classification \cite{bobda2014distributed}. The results of implementation are shown on Figure  \ref{result1} and \ref{result2} as snapshot from the video processing. 

Having completed the first components of our control infrastructure, namely extracting knowledge from images to supply waiting queues in real-time, we first perform some analysis of performance before devoting the rest of the paper to the remaining modules of our architecture. 

\subsection{Performance Analysis}
We measured the computation time of our model as illustrated in Table \ref{tab:perf}. We tested our model as a single-core implementation on a 64-bit x86 processor with a clock frequency of 3.60GHz. The total computation time does not include the calibration time since this will be only performed once only at the beginning of the computation. For calibration and testing purposes, We used a 4-minutes video clip of a 4 -way transaction as input. The results suggest that the model can achieve close to 4 frames per second. This can be improved by trading the detection model with a lighter one. We were able to achieve 13 frames per second by using the YOLO-V3-tiny model to detect vehicles. However, this reduces the accuracy of the model as well.

\begin{table}[H]
\scriptsize
\centering
 \caption{Performance Analysis}
 \label{tab:perf}
\begin{tabular}{@{}lc@{}}
\toprule
\multicolumn{1}{c}{\textbf{-}} & \multicolumn{1}{l}{\textbf{Computation time (ms)}} \\ \midrule
\textbf{Detection time} & 224ms \\
\textbf{Calculating traffic density} & 40ms \\
\textbf{Total Computation time} & 262ms \\
\textbf{Frames per second} & 3.82fps \\ \bottomrule
\end{tabular}
\end{table}


%% file: section4-WaitingQueue.tex
\section{Waiting Queue}\label{sec:wait}
Our algorithm operates on waiting queues, which are updated using knowledge gained from the image processing module. This section provides details of the various waiting queues (WQ) used in our model.

A waiting queue represents vehicles in a given lane. We consider two types of waiting queues in the target system: Entry queues and exit queues:

\subsection{ Exit Queue}
An Exit Queue noted $Q_O$ or output queue at the intersection can be in one of two states : 
\begin{itemize}
\item[$\bullet$]  \textbf{Open} Meaning this queue  can accommodate vehicles.  
\item[$\bullet$] \textbf{Closed} Meaning this queue can not accommodate new vehicles. 
\end{itemize} 
For a queue in open state, we find the free space by using the density over the lane space.. This information is  important for finding  the  maximum number of vehicles that the queue can accommodate. 
\subsection{ Entry Queue}
An Entry queue $Q_I$ or Input queue at the  intersection has three  possible states : 
\begin{itemize}
\item[$\bullet$]$\{A\}$: \emph{Active} means that the signal light is \textbf{green} (G) for this queue.
\item[$\bullet$]$\{WA\}$: \emph{Waiting Active} meaning that the signal light should be \textbf{green}  but  for some reason($Q_O$ is closed or important vehicle has been detected) it has been blocked and the  light remains red.
\item[$\bullet$] $ \{IA \} $: \emph{Inactive} meaning that the signal light is \textbf{red} (R) for this queue. 
\end{itemize}

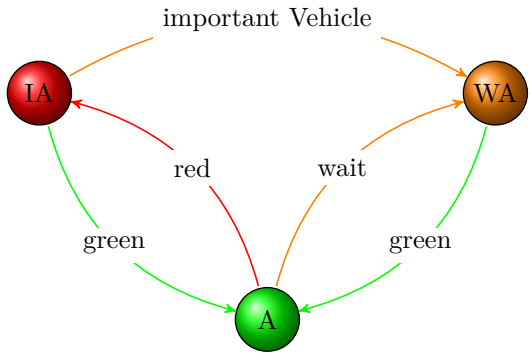
\begin{figure}[H]
\centering
\begin{tikzpicture}
\SetUpEdge[lw= 1.5pt]
\GraphInit[vstyle=Shade]
\begin{scope}[VertexStyle/.append style={shape=circle, shading=ball, ball color=red}]
\Vertex[x=0 ,y=0]{IA}
\end{scope}
\begin{scope}[VertexStyle/.append style={shape=circle, shading=ball, ball color=green}]
\Vertex[x=3 ,y=-3]{A}
\end{scope}
\begin{scope}[VertexStyle/.append style={shape=circle, shading=ball, ball color=orange}]
\Vertex[x=6 ,y=0]{WA}
\end{scope}

\tikzset{EdgeStyle/.style={draw=red, post, bend right}}
\Edge[label=$\text{red}$](A)(IA)

\tikzset{EdgeStyle/.style={draw=orange, post, bend left}}
\Edge[label=$\text{wait}$](A)(WA)
\Edge[label=$\text{important Vehicle}$](IA)(WA)
\tikzset{EdgeStyle/.style={draw=green, post, bend left}}
\Edge[label=$\text{green}$](WA)(A)

\tikzset{EdgeStyle/.style={draw=green, post, bend right}}
\Edge[label=$\text{green}$](IA)(A)

\end{tikzpicture}
\caption{Simple state diagram of queue.} 
\label{DiagramBig}
\end{figure}
\begin{figure}[!h]
 \captionsetup{justification=centering}
    \begin{subfigure}{.48\textwidth}
       \centering
        \captionsetup{justification=centering}
        \includegraphics[width=0.7\linewidth]{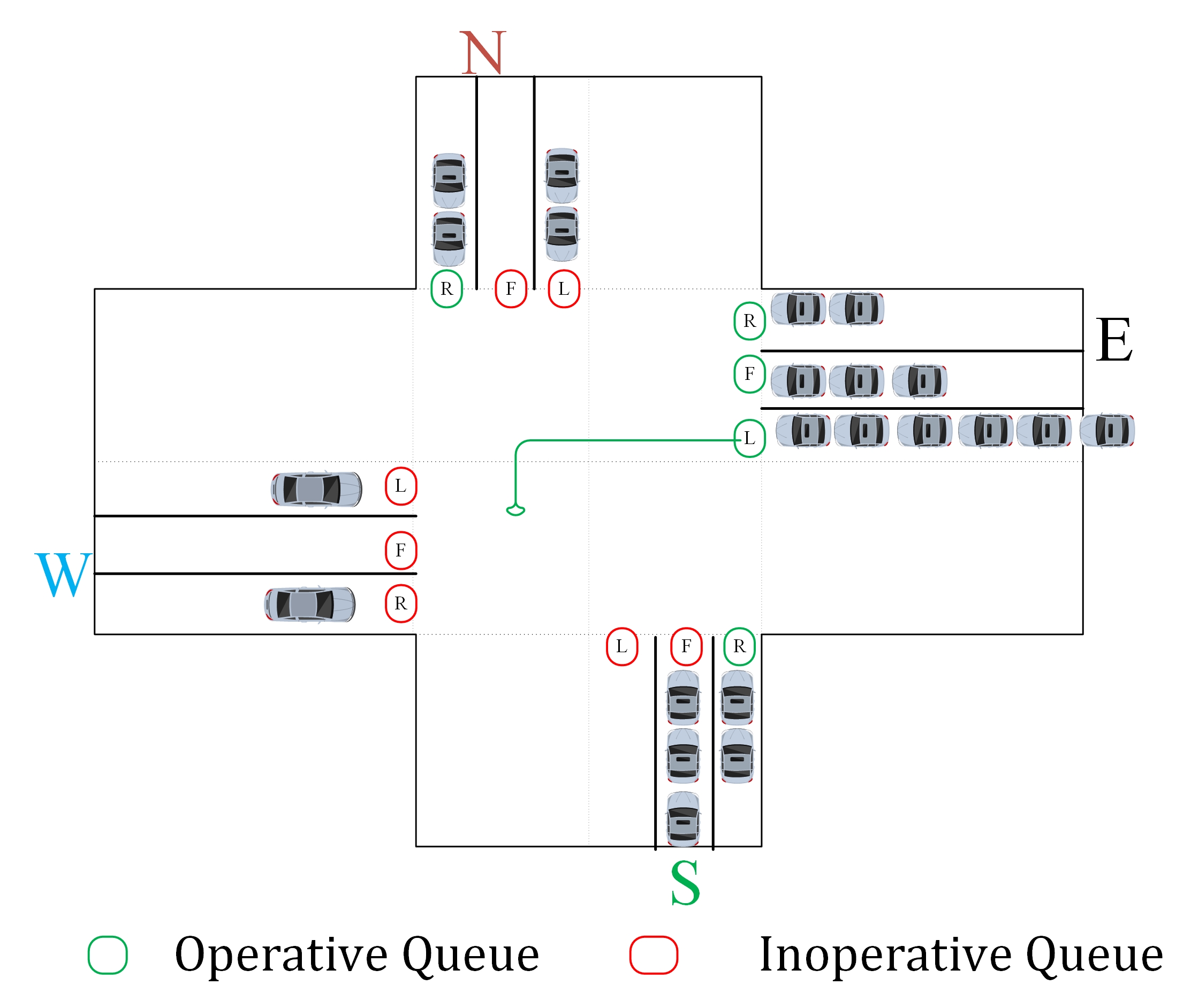}
        \caption{Our approach consists of setting the green light to queue with highest number of car. At time $ t$:  $\{NL,WR, WL\}$ are waiting.}
        \label{exa1}
\end{subfigure}
    \begin{subfigure}{.5\textwidth}
        \centering
         \captionsetup{justification=centering}
        \includegraphics[width=0.68\linewidth]{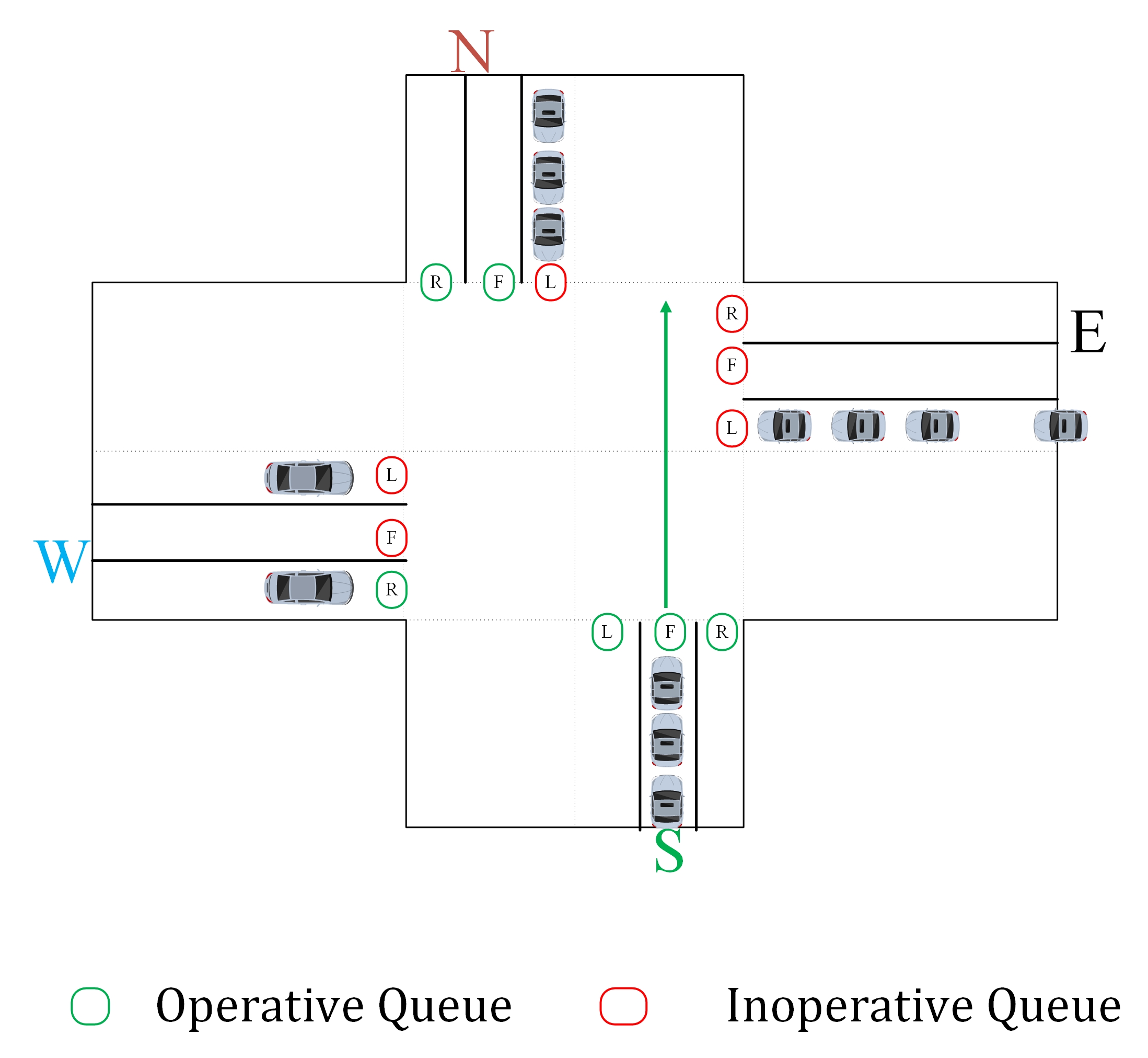}
        \caption{Apply same principle as shown on figure  \ref{exa1}.  At time $t+x$ :  $\{NL,WR\}$ are still waiting and next queue to be opened is $\{EL\}$. So, $\{NL,WR\}$ might wait forever. }
        \label{exa2}
        \end{subfigure}
        \caption{Dead wait of  low traffic queue along heavy traffic queue for simple state diagram queue  modeling.}
\end{figure}

 
Figure \ref{DiagramBig} shows the states a queue can be in at any time. The controller algorithm chooses the queue with the highest number of vehicles to move from state $\{IA\}$ to $\{A\}$ and the state $\{WA\}$ precedes $\{IA\}$. However, the logic to select the waiting queue with a large number of vehicles has some problems, namely long waiting time in queue with low traffic. 
 Figure \ref{exa1} shows that at time $t$, $\{EL\}$ has the highest number of cars in it's queue so the algorithm turns this light green, at the same time $\{EF, ER,NR,SR \}$ can be open without conflict. In Figure \ref{exa2} a time $t+x$, $\{ SF\}$ is the queue with most of the cars, so this turns green. Observe that $\{ EL\}$ is filling up with vehicles and will be opened next time, while queues such as $\{ NL\}$ and $\{ WL\}$ with low traffic are still waiting and may wait forever. \par

\begin{figure}[!h]
\centering
\begin{tikzpicture}
\SetUpEdge[lw= 1.5pt]
\GraphInit[vstyle=Shade]
\begin{scope}[VertexStyle/.append style={shape=circle, shading=ball, ball color=red}]
\Vertex[x=0 ,y=0]{0}
\Vertex[x=2 ,y=0]{1}
\Vertex[x=4 ,y=0]{2}
\Vertex[x=6 ,y=0]{3}
\Vertex[x=5 ,y=-2]{4}
\end{scope}
\begin{scope}[VertexStyle/.append style={shape=circle, shading=ball, ball color=green}]
\Vertex[x=1 ,y=-2]{A}
\end{scope}
\begin{scope}[VertexStyle/.append style={shape=circle, shading=ball, ball color=orange}]
\Vertex[x=3 ,y=-4]{WA}
\end{scope}
\tikzset{LabelStyle/.style={draw,fill = white,text = red,post}}
\tikzset{EdgeStyle/.style={draw=red, post, bend left}}
\Edge[label=$Y_i(t)$](0)(1)
\Edge[label=$Y_i(t)$](1)(2)
\Edge[label=$Y_i(t)$](2)(3)
\Edge[label=$Y_i(t)$](3)(4)

\tikzset{LabelStyle/.style={draw,fill = white,text = orange,post}}
\tikzset{EdgeStyle/.style={draw=orange, post}}
\Edge[label=$E$](0)(4)
\Edge[label=$E$](1)(4)
\Edge[label=$E$](2)(4)
\tikzset{EdgeStyle/.style={draw=orange, post, bend right}}
\Edge[label=$E$](3)(4)

\tikzset{EdgeStyle/.style={draw=red, post, bend right}}
\Edge[](A)(0)

\tikzset{EdgeStyle/.style={draw=orange, post, bend left}}
\Edge[](A)(WA)
\Edge[label=$E$](4)(WA)
\tikzset{EdgeStyle/.style={draw=green, post, bend left}}
\Edge[](WA)(A)

\tikzset{LabelStyle/.style={draw,fill = white,text = green,post}}
\tikzset{EdgeStyle/.style={draw=green, post, bend right}}
\Edge[label=$X_t(t)$](0)(A)
\tikzset{EdgeStyle/.style={draw=green, post}}
\Edge[label=$X_i(t)$](1)(A)
\Edge[label=$X_i(t)$](4)(A)
\tikzset{EdgeStyle/.style={draw=green, post, bend left}}
\Edge[label=$X_i(t)$](2)(A)
\Edge[label=$X_i(t)$](3)(A)
		\Edge[label=$X_i(t)$](WA)(A)

\end{tikzpicture}
\caption{ Extend state diagram of a queue.} \label{Diagram-state}
\end{figure}
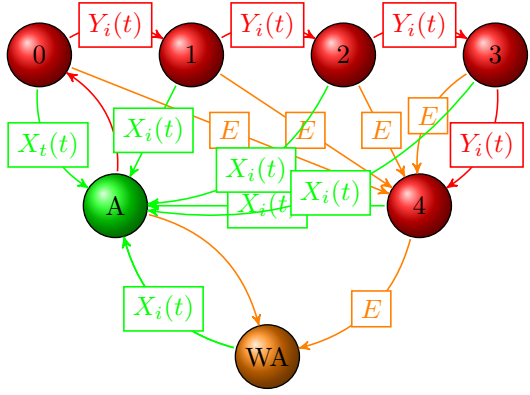

We have introduced the notion of internal state $IA \in \{0,1,2,3,4 \}$ (the light is still red for those states) to take into account the fact that a queue with low traffic can at certain time have the highest priority. As describe in Figure \ref{Diagram-state} where
\begin{itemize}
\item E is a transition caused by the presence of an emergency vehicle.
\item $Y_i(t)$ is elapsed time for a queue to change his internal state during the cycle $i$.
\item $X_i(t)$ is the time allocated to empty the queue during a cycle $i$. 
\end{itemize}
A cycle is when a queue made a complete turn through state  $\{A\}$. 
 An example of a possible cycle can be $\{ 0 \rightarrow 1 \rightarrow 2 \rightarrow A \rightarrow 0\}$, or$\{ 0 \rightarrow 1 \rightarrow 2 \rightarrow 3 \rightarrow WA \rightarrow  A \rightarrow 0\}$. Hence $Y_i(t)$ is computing each cycle and is constant for a time slot $t$. The figure \ref{Diagram-state} shows the difference transitions and state of a waiting queue $Q_I$. 

The set $\{0,1,2,3,4 \}$ are  the states that a waiting queue $Q_I$ goes trough cyclically whenever transition conditions are fulfilled. Only the controller or the master can set the status of the queue as $\{A\}$ and afterwards to zero to mark the end of a cycle. \par
Let's consider this cycle as the first running time. $\{ 0 \rightarrow 1   \rightarrow 4 \rightarrow  A \rightarrow 0\}$.  The queue starts in state $\{ 0\}$, we assume that, the queue is not empty (id the queue is empty, the state remains in zero). After $Y_1(t)$ time has elapsed (index 1 means cycle number 1) the queue moves to state $\{ 1\}$. The presence of an emergency vehicle upgrades it to state $\{ 4\}$, and after some time it becomes the only queue in state $\{ 4\}$, and it moves to state $\{ A\}$. After the time taken to empty the queue noted $X_1(t)$, it returns to state $\{ 0\}$ and $Y_2(t)$ is computing to mark the end of the cycle.  \par 
Let $I^{a}_i(t)$ be the internal state of queue $\{a\}$ during the cycle $i$, the equation is given by:
\begin{equation}
\tiny
I^{a}_{i}(t) = \begin{cases}
I^{a}_{i}(t-1),  \text{ if } I^{a}_{i}(t-1) < 0 \text{ or } I^{a}_{i}(t-1) > I_{max} \\
\min\{I^{a}_{i}(t-1)+1_{[Y^{a}_{i} ~\text{elapsed} ~ \&~ d^{a}_I(t)\neq 0 ]},I_{max}\},  \text{ Otherwise}
\end{cases}
 \label{eqInternalState}
\end{equation}  
Where $I \in \{ 0,1,\cdots,I_{max}\}$ , in particular  $I_{max}=4$, $1_{[X]}$ is the indicator function that takes the value $1$ if $X$ is true and $0$ otherwise, and $d^{a}_I(t)$ the current density of the entry queue $\{ a\}$.  \par 

There are 12 queues in our target system: $\{WR,WF$ $,WL,ER,EF,EL,NR,NF,NL,SR,SF,SL\}$, each of 
which has an internal state. This internal state is relayed back to the main controller, where the algorithm sets the queue with the highest state to have a green light. If this is not possible, it selects the next highest state queue. After the queue has been emptied, the controller then reassesses which queue should be given priority.

\subsection{Calculating $X_i(t) $ : time needed to empty a queue}

\par $X_i(t)$ is the time needed to empty the  $Q_I$ at time slot $t$ during a cycle $i$. This time depends on the exit  and the entry queue. \par
\begin{figure}[!h]
       \centerline{
        \includegraphics[width=2.5in]{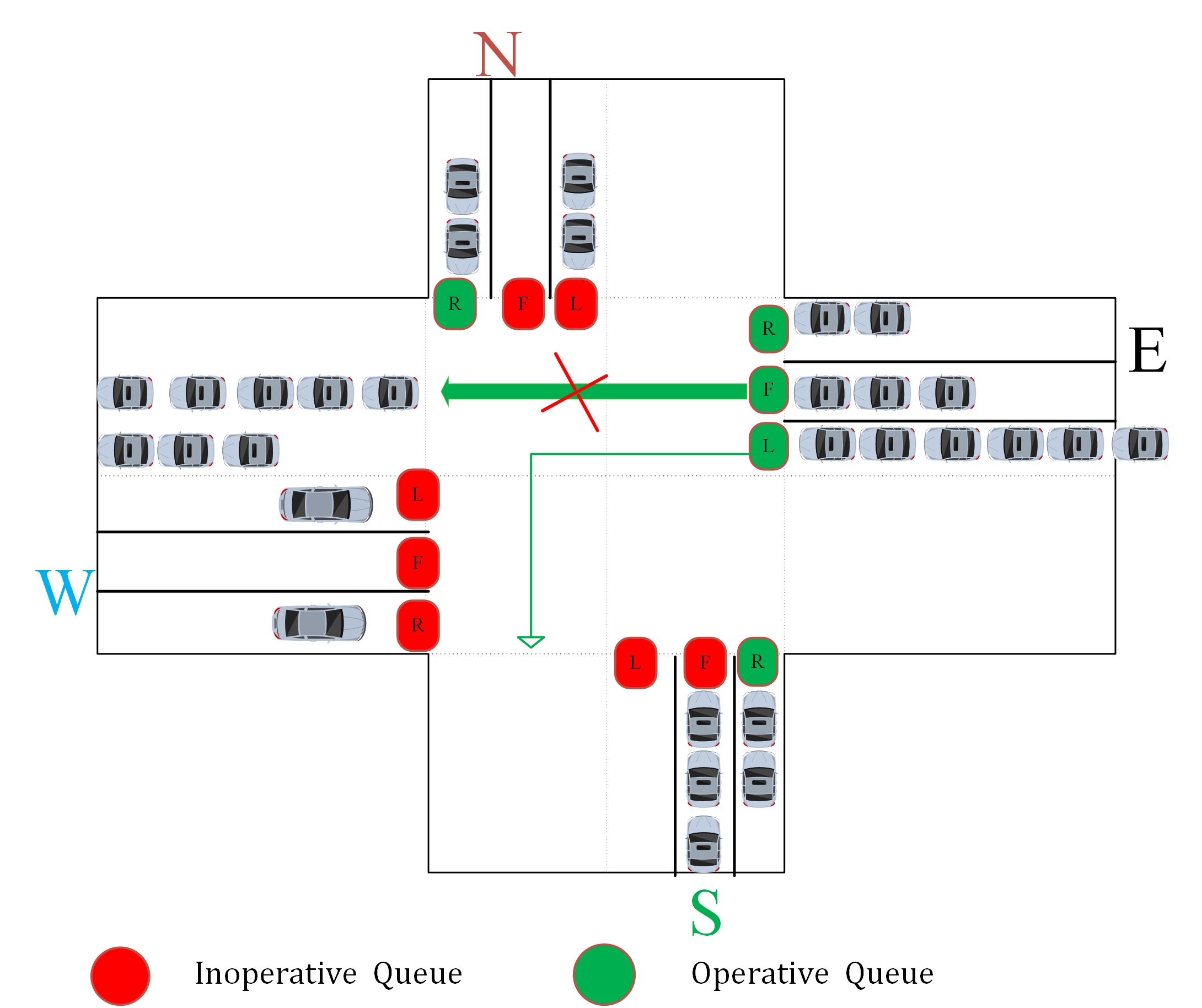}}
        \caption{Case of a queue EF with green light but with exit queue that cannot accommodate new vehicle.}
        \label{Counter-Example}
\end{figure}

There are  situations where  the exit queue is not ready to accommodate vehicles but the light is green  for the entry queue. On the figure \ref{Counter-Example}, The queue $\{EL\}$ has the highest number of vehicles and has been selected, then  we find the queue in green  can operate without conflict, especially the arrival queue $\{ EF\}$ but the departure queue is not ready to accommodate vehicles. 
\par 
Hence, we take into account all these parameters and find out that the formula for computing the time $X^{a}_{i}(t)$ needed to empty a queue $\{a\}$ during a cycle $i \in \mathbb{N}$ at time slot $t$ is:  
\begin{equation}
X^{a}_{i}(t) = \frac{\min\{(1-d^{a}_{O}(t))L^{a}_{O},d^{a}_{I}(t).L^{a}_{I}\}}{V} .\label{Xtime}
\end{equation}   
Where $d^{a}_{I}$ and $L^{a}_{I}$ are the current density of the vehicle on input traffic queue $\{ a\}$ and the length of the queue $\{ a\}$, $d^{a}_{O}$ and $L^{a}_{O}$ are the current density on the output traffic queue(exit queue) and the length of the queue, and  $V$ is the velocity to cross the crossroad. 

\begin{itemize}
\item $(1-d^{a}_{O}(t))L^{a}_{O}$ is the proportion of space  available to accommodate vehicles on exit (output) queue. 
\item $d^{a}_{I}(t).L^{a}_{I}$ is the proportion of space occupied by vehicles on the entry (Input) queue.
\end{itemize}

The function minimum guaranty the time needed to empty a queue is enough in order to avoid vehicles to remain in middle of the intersection after that time elapsed.  \par 
Note that, $X^{a}_{i}(t)=0 $ implies either  $d^{a}_{I}=0$  meaning there is not vehicles on the input queue, or  $d^{a}_{O}=1$ meaning the Output queue is full and can not accommodate vehicles.  

\subsection{Calculating $Y_i(t)$}
\par Let $Y^{a}_{i}$ be the time needed for a queue $\{a\}$ to change internal state. $Y^{a}_{i}$ can be written as a function of the time needed to empty the queue: 
\begin{equation}
Y^{a}_{i} = f(X^{a}_{i-1}) \label{Ytime}.
\end{equation}  

Our goal is to find an appropriate function $f$ that minimizes the time needed to empty the queue. If this time is high, this queue will be assume to have heavy traffic, and its internal state should change quickly. That is, if $X^{a}_{i-1}$ is large, then $Y^{a}_{i}$ should be small, and vice versa.   
\begin{figure}[!h]
        \centering
         \captionsetup{justification=centering}
        \includegraphics[width=0.4\textwidth]{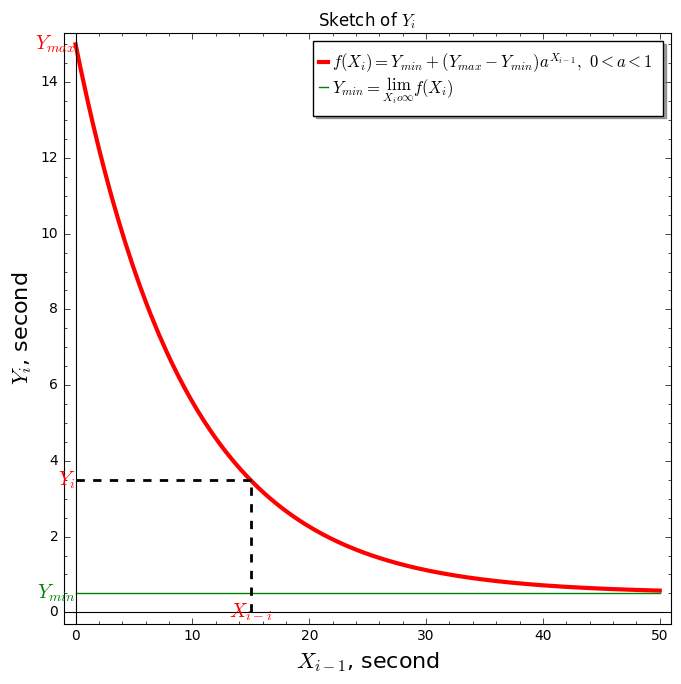}
        \caption{Plot of a possible choice for $Y^{a}_{i} = f(X^{a}_{i-1})$ }
        \label{sketchY}
\end{figure}
Thus equation \ref{Ytime} becomes : 
\begin{equation}
Y^{a}_{i} = Y_{min}+(Y_{max}-Y_{min})a^{X_{i-1}} ~~ \text{where} ~~ 0<a<1
 \label{YtimeFinal}
\end{equation} 
The figure \ref{sketchY}  was obtained with $a=\frac{4.5}{5}, Y_{min}=0.5,$ and $Y_{max}=15$.  \par
Here we will make an assumption that the model \ref{sketchY} is appropriate. Our goal is to vary the parameters $\{ a, Y_{min},Y_{max} \}$ in order to minimize the average waiting time. \par 

Note that if a queue has only one car, it's internal state will still change, but the time taken to change will be longer than that of a queue with many cars. This way both the number of cars and the time spent are taken into account when assigning an internal state.  \par 

\begin{figure}[!h]
\centering
 \captionsetup{justification=centering}
\includegraphics[width=0.3\textwidth]{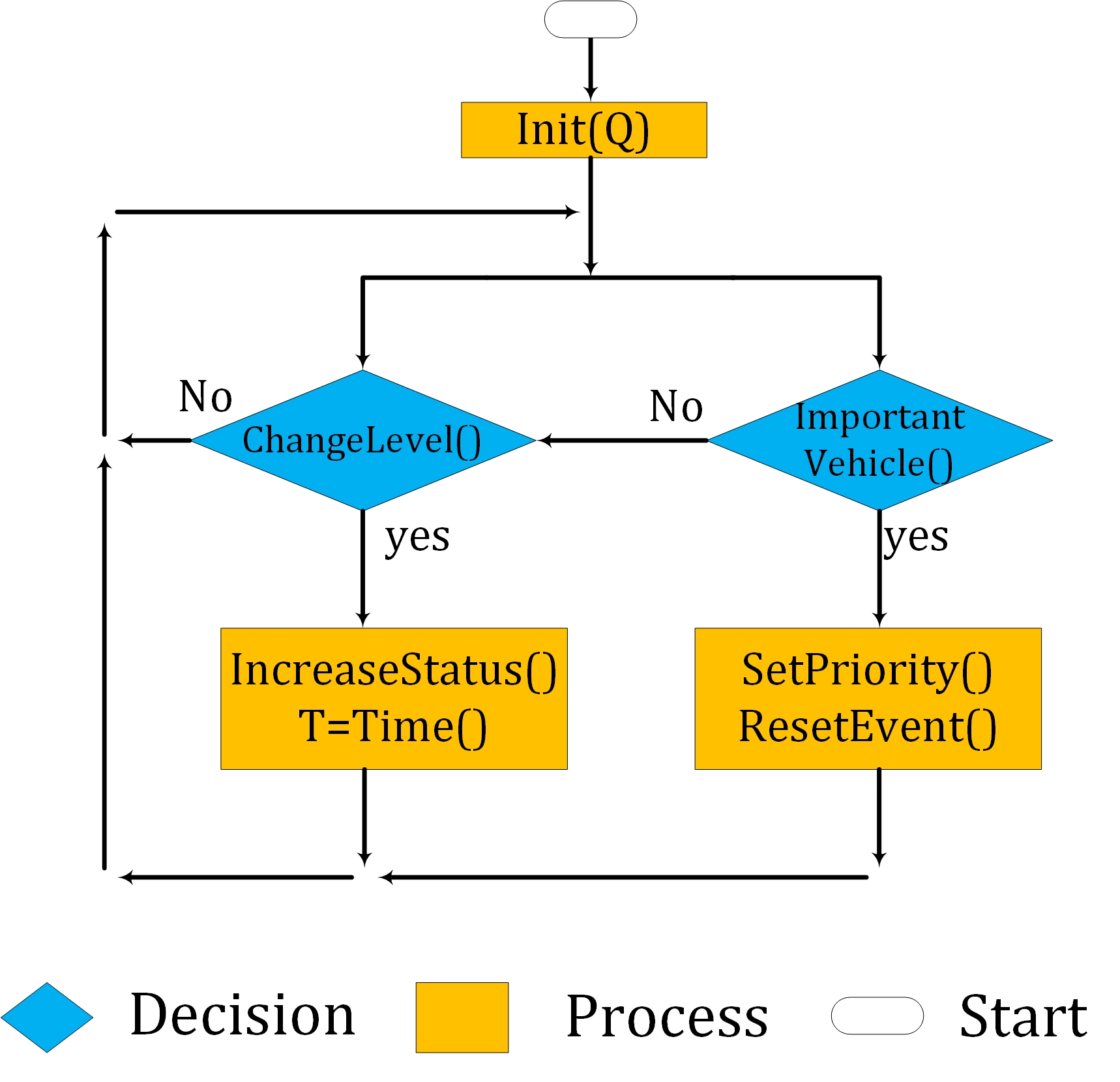}
\caption{Flowchart queue insight.}
\label{flowChartQueue}
\end{figure}

The complete algorithm is described as flowchart on the figure \ref{flowChartQueue}. Note  that, each queue is supposed to run independently this algorithm.  

\begin{table}[H]
\tiny
\centering
\caption{Sub functions used on flowchart Figure \ref{flowChartQueue}.}
\begin{tabular}{@{}ll@{}}
\toprule
\multicolumn{2}{c}{\textbf{Sub-process}} \\ \midrule
\textbf{Init()} & $Y_i=f(X_{i-1})$;Status=0; Event=$False$; T=Time(). \\
\textbf{ChangeLevel()} & Check if $Y_i$ time elapsed and  $d \neq 0$ and status $\neq \{A, WA \}$. \\
\textbf{IncreaseStatus()} & Status = $I_i(t+1)$, get it from equation \ref{eqInternalState} \\
\textbf{VehicleImportant() }& Event= ( emergency vehicle detected) ? True : False \\
\textbf{SetPriority()} & Set Status = (Status \textless 4) ? 4 : WA. \\
ResetEvent() & Event=False \\ \bottomrule
\end{tabular}
\label{subfunctioon}
\end{table}

\begin{table}[H]
\tiny
\centering
\caption{Variables  used  on flowchart figure \ref{flowChartQueue}.}
\begin{tabular}{@{}ll@{}}
\toprule
\multicolumn{2}{c}{\textbf{Variables}} \\ \midrule
\textbf{T} & Record the  time when increase Status occurs. \\
\textbf{Status} & Current Internal state of a queue, $\in \{0,...,T_{max}, WA, A \}$ \\
\textbf{Y\_i} & Time elapsed for changing the internal state during current cycle. \\
\textbf{d} & Current density of the queue. \\
\textbf{X} & Time needed to empty queue during the previous cycle \\ \bottomrule
\end{tabular}

\label{variableused}
\end{table}

%% file: section5-DistributedAlgo.tex
\section{Distributed Control Algorithm}\label{sec:distributed}

In this section we describe the distributed algorithm for traffic light, based on the conflict matrix \ref{MatrixCon}.

\subsection{Modeling}
We have described the different states of a queue in section  \ref{Diagram-state}, taking into account the density of traffic and the time elapsed. The road code was modeled through the conflict table \ref{MatrixCon} where entries in the diagonal were considered the illegal case. In the rest of this article, this table will be seen as a square symmetric matrix where indexes are queuing labels and the diagonal entries will hold the state of the corresponding queue. This matrix named here $M$ contains the configuration of the intersection and the current state of the system. \par Thus the  square  matrix $M$ corresponds to the states of the queue and being constructed based on the following formula: 
\begin{equation}
\scriptsize
M_{[a,b]}(t) =\begin{cases}
I^{a}_{i}(t) ,  & \text{ if } a = b \\
\otimes, &  \text{if queue } \{a\} \text{is in conflict with queue. }  \{b\} \\
0,   & \text{Otherwise. }
\end{cases} 
\label{eqModelingMatrix}
\end{equation}
where $I^{a}_{i}(t)$ is the current state of queue $\{ a\}$  from \ref{eqInternalState}. Note that, from the equation \eqref{eqModelingMatrix} are follow :
\begin{equation*}
M_{[a,b]} = M_{[b,a]} \text{ and } M_{[a,a]} = I^{a}_{i}
\end{equation*}
To solve this, we introduce a numerical values for the active state $A:-1$. and similarly for the waiting active state, we assign it a value of $I_{max}+1$ which is $5$ in particular.
\begin{equation*}
\{A\} \leftarrow -1 \text{ and }\{WA\} \leftarrow  5  
\end{equation*}
$\{A\}$ and $\{WA\}$ are respectively the status Active and Waiting Active from our state diagram of Figure \ref{DiagramBig}. The matrix $M$ of equation \ref{algoMatrix} represents the state of the system at a specific time where the diagonal is the only dynamic part.  The queues $\{ ER,NR\}$ in red are operative, thus their internal state is set as  $\{A\}$ meaning $M_{[ER,ER]} = - 1 $, and the next queue to open is $\{SF\}$ because of its highest internal state obtaining with following formula $ $ $\argmax\{diagonal(M)\}=\arg\{M_{[i,i]}=3\}=\{SF\}$.


\begin{equation}
\scriptsize
\centering
M=\begin{blockarray}{cccccccc}
WR & \cdots & {\color{red} ER} & \cdots & {\color{red} NR} & \cdots & {\color{blue} SF}   \\
\begin{block}{(ccccccc)c}
{\color{yellow} 2}	&  \cdots	&0 &	\cdots &0 & \cdots &0 & WR	\\
\vdots & \ddots & \vdots & \ddots & \vdots & \ddots & \vdots & \vdots \\
0 & \cdots & {\color{red} -1} & \cdots & 0 &\cdots & \otimes & {\color{red} ER} \\
\vdots & \ddots & \vdots & \ddots & \vdots & \ddots & \vdots & \vdots \\
0	&  \cdots	&0 &	\cdots &{\color{red} -1} & \cdots &0  & {\color{red} NR}	\\
\vdots & \ddots & \vdots & \ddots & \vdots & \ddots & \vdots & \vdots \\
0 & \cdots & \otimes & \cdots & 0 &\cdots & {\color{blue} 3} & {\color{blue} SF} \\
\end{block}
\end{blockarray}
\label{algoMatrix}
\end{equation}
This model with the matrix gives us a flexibility in case of modification of the road code at the intersection, it will be enough for us to modify the static part of the matrix by performing an operation to exchange between $\{0\}$ and $\{\otimes \}$. These operates can be extend in  remotely control the intersection for a smart city. \par

\subsection{Algorithm for updating the matrix $M$}
Since each queue has his internal state as described in section \ref{flowChartQueue}, it is enough to set their internal state in dynamic part of the matrix $M$.  \par
\begin{algorithm}[!ht]
\scriptsize
\SetAlgoLined
 \KwData{set of the queue }
\KwResult{ Matrix $M$  }
\For{$q \in set Of Queue$}{
\tcp{Compute the internal state of the queue from the  flowchart \ref{flowChartQueue}}
     Compute  $I^q$ \;
     \tcp{Update the diagonal of the matrix $M$}
     $M[q,q] \leftarrow I^q $\;
}
 \caption{Update the matrix $M$ }
 \label{UpdateMatrix}
\end{algorithm}

\subsection{ Notations and Sub-functions }
Now that we have digitized any intersection via our Matrix $M$, before writing the distributed algorithm, we are going to show variables and sub-functions used in the algorithm. 
\begin{table}[H]
\tiny
\caption{Notations used in algorithm \ref{AlgoDistribted}.}
\centering
\begin{tabular}{@{}ll@{}}
\toprule
\multicolumn{1}{c}{\textbf{Variables}} & \multicolumn{1}{c}{\textbf{Meaning}} \\ \midrule
$Q$ & \begin{tabular}[c]{@{}l@{}}Set  of all queue \\  i.e $Q=\{WR,WF, WL, ER ,EF ,EL ,NR, NF ,NL, SR ,SF ,SL\}$\end{tabular} \\
\textbf{$M$} & Square Matrix of size $\| Q\|\times\| Q\|$ \\
\textbf{ListOpen} & Set of queue to open \\
\textbf{RunList} & Set of queue in process \\
S & Set of forbidden queue \\ 
$U_a$ & set a queue without conflict with queue $\{a\}$ \\ \bottomrule
\end{tabular}
\end{table}
For the \textbf{Sub functions}, we have two. $ Sort(L)$ : which  Sorts in ascending order the items in the list $L$, and $Open(a)$.
\begin{equation*}
\scriptsize
Open(a) =\begin{cases}
\text{ True} , & if X^a(t) \neq 0 \text{ we obtain } X^a \text{ from } equation \ref{Xtime}  \\ 
& X^a(t) \text{: time needed to empty the queue } \{a\}  \\
\text{False}, &  \text{Otherwise.}
\end{cases} 
\label{eqModelingMatrix}
\end{equation*}
Basically, the sub-function $open(a)$ is a Boolean function that returns true if the time needed to empty the queue is different to zero and false otherwise. Consider a scenario where an entry queue is full and the exit queue is also full. Then at a certain time, we realize that the queue has the highest priority. If we set a queue to green, the vehicles will stand in the middle of the intersection and that will cause traffic congestion. In order to avoid that, the time $X^a_i$  needed to empty a queue is taking into account this scenario.

The distributed algorithm \ref{AlgoDistribted} is based on the matrix $M$ essentially the diagonal of the matrix. It operates according to the principle that the queue having a high priority level (internal state) will be selected, and this algorithm is supposed to be executed by an irregular interval of time.

\begin{algorithm}[!h]
\scriptsize
 \caption{Distributed Algorithm.}
\SetAlgoLined
 \KwData{Matrix $M$ }
\KwResult{ Return the set of queue to open $ListOpen$  }
\tcp{initialization}
 $ListOpen \leftarrow \oslash $ \;
 \tcp{get  set a queue in process}
$RunList \leftarrow \arg \{{diag(M)<0} \}$ \;
$S \leftarrow RunList$ \;
\tcp{find set of queue which is in conflict with queues in Process}
  \For{$q \in RunList$}{
     $p \leftarrow \smash{\displaystyle\arg_{i \in Q}} \{M[q,i]=\otimes \}$ \;
    $ S \leftarrow S \cup \{p\}$ \;
}
 $listOrder \leftarrow \smash{\displaystyle\arg_{i \in Q\backslash S}} \{Sort(M[i,i]) \} $ \;
 \tcp{find exactly one queue without constraint to open}
  \For{$q \in ListOrder$}{
          $S \leftarrow S \cup \{q\}$ \;
    \eIf{ $Open(q)$}{
        $M[q,q] \leftarrow WA=5$ \;
        \tcp{ Take the next queue with highest level}
        $continue$\;
    }{
    $ ListOpen \leftarrow \{q \}$ \;
    $break$ \;
    }
}

\eIf{ $ ListOpen = \oslash $}{
\tcp{ End the program if there no queue to open}
      $ exit()$ \;
}{
    \tcp{ find the list of queue without conflict with $q \in ListOpen$}
    $U_q \leftarrow \smash{\displaystyle\arg_{i \in Q\backslash S}} \{M[q,i])\neq \otimes \}$ \;
    \tcp{Find  the set of queue to open in parallel with $q$ without conflict}
    \While {$U_q \neq \oslash $}{
    \tcp{Among those queues, chose the one with  highest internal state}
    $p \leftarrow \smash{\displaystyle\argmax_{i \in U_q}} \{M[i,i] \}$ \;
    \tcp{}
    \If{$Open(p)$}{
    \tcp{add $p$ in the list of queue to open}
     $ ListOpen \leftarrow ListOpen \cup \{p\}$ \;
    }
    \tcp{remove $p$ from $U_q$}
    $ U_q \leftarrow U_q \backslash \{p\}$ \;
    \tcp{ find the list of queue without conflict with $p$}
   $ U_p \leftarrow \smash{\displaystyle\arg_{i \in U_q}} \{M[p,i] \neq \otimes \}$ \;
   $ U_q \leftarrow U_q \cap U_p$
   
    }
    
}
 \label{AlgoDistribted}
\end{algorithm}

%% file: section5_Evaluation.tex
\section{Simulation }\label{sec:simulation}
This section evaluates our algorithm using a combination of image processing and the SUMO (Simulation of Urban MObility,\cite{SUMO2012}) framework. SUMO is an open-source, discrete-time, continuous space, microscopic simulator entirely coded in C++ to model traffic flow as well as supporting tools, mainly for network import and demand modeling. In short, it allows placing sensors and retrieving real-time traffic data using tools TraCI (\textbf{Tra}ffic \textbf{C}ontrol \textbf{I}nterface ) and to act on the behavior of the Traffic Light Control on-line. Our distributed algorithm is based on an intersection modeled from the city Amiens in France at rush hour, between 8 AM and 9 AM, which according to  Sebastien Faye and al. \cite{faye2012distributed2} statistically have in SUMO new vehicles arrival-rate of $\lambda = 0.8$ per second on the intersection. Relying on this statistical assumption, our simulation was conducted on an intersection with 4 directions and 12 possible movements: from each direction, a vehicle could go straight, turn left or turn right. In addition to that, we injected emergency vehicles with an arrival rate of $\lambda = 0.025 $ per second in order to evaluate the impact of such vehicles on our distributed algorithm. Each simulation ran for 3600 program steps representing the 3600s. The results presented below are the average waiting time computed on the 3000 first vehicles that left the intersection. This allows us to evaluate our algorithm with a realistic traffic. Note that, the TraCI tool allows us to gather traffic data but in the real deployment of our proposed solution, those traffic data are provided by the distributed smart-cameras installed at the intersection. 
\begin{figure}[!ht]
        \centering
         \captionsetup{justification=centering}
        \includegraphics[width=0.5\textwidth]{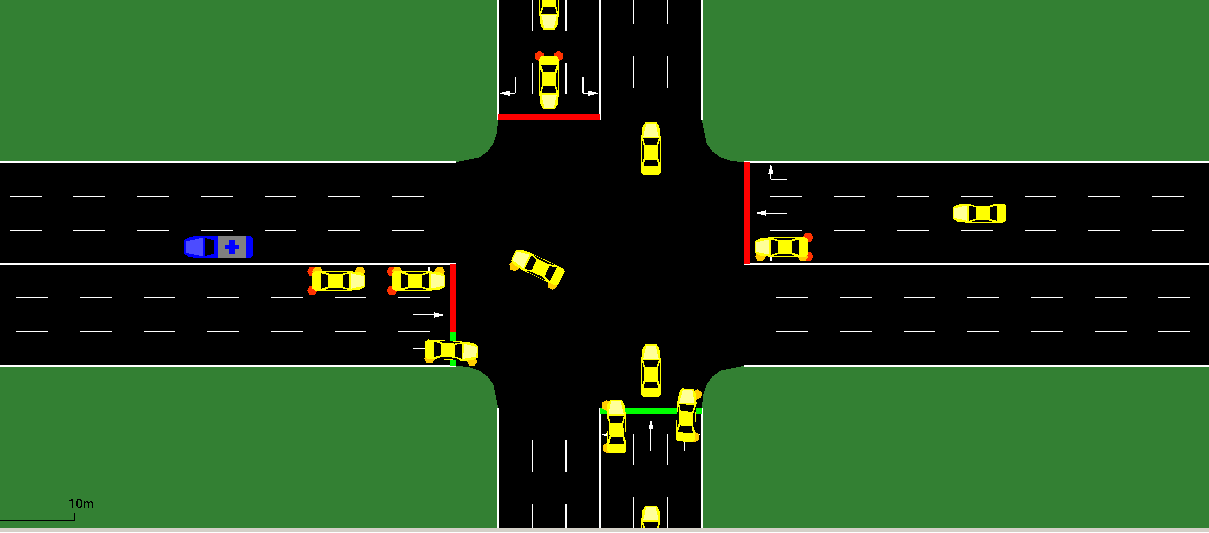}
        \caption{Capture of simulation after 150 steps(150 seconds), Emergency vehicle(EV) in blue and Classic vehicles(CV) in yellow. }
        \label{fig :simulation}
\end{figure}
\subsection{Parameters}
The green light limit time $T_{max}$ are expected  to have a strong  influence  on the performance result. In case $T_{max}< X_i$ (define time needed to empty queue $i$  see equation \ref{Xtime}), the green light queue will be  $T_{max}$. Afterwards, the internal state  of  the queue will not  be  set  to zero but proportionally  to the  ratio between $T_{max} $  and $ X_i$. We chose to evaluate the algorithm  with two scenarios :(1) $S_1$  when  taking into account the priority of the emergency vehicles; (2) $S_2$ without considering priority on any vehicles. For both scenarios, we output the average waiting time of emergency vehicles(EV), classic vehicles (CV) and both vehicles (AV). These simulation is conducted  for $T_{max} \in [15,90]$ in steps of $5$ seconds.
\begin{figure}[!h]
\begin{subfigure}{.5\textwidth}
  \centering
        \centering
        \captionsetup{justification=centering}
        \includegraphics[width=0.8\textwidth,scale=0.8]{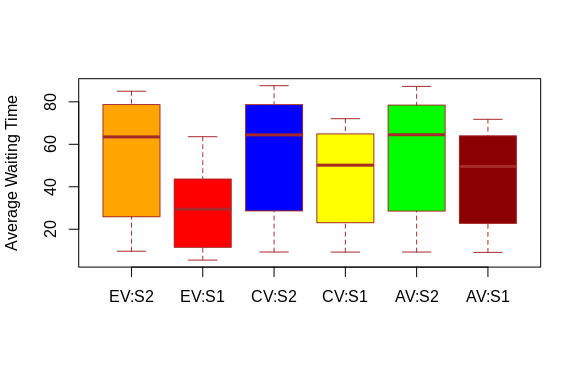}
        \caption{Multiple average waiting time for scenarios  $S_1$ \\ and $S_2$ with collision forbidden. }
        \label{fig : awtboxplot}
\end{subfigure}
\begin{subfigure}{.5\textwidth}
        \centering
         \captionsetup{justification=centering}
        \includegraphics[width=0.8\textwidth,scale=0.8]{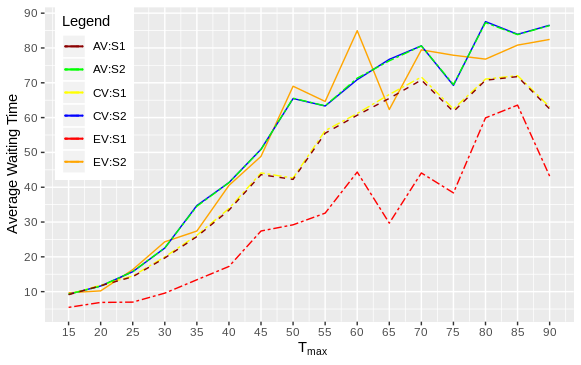}
        \caption{(Vehicle type, scenario), $T_{max}  $  influence on average waiting time with collision forbidden. }
        \label{fig : awtTmax}
\end{subfigure}
\caption{Performance results.}
\label{fig:AWT_Result}
\end{figure}

 Figures \ref{fig : awtboxplot} and \ref{fig : awtTmax} present the average waiting time at an intersection for different class of vehicles with and without emergency vehicles. We can first  notice that, the scenario $S_1$ allows reaching the better average waiting time especially the emergency vehicles (EV) which is the target class of vehicle for this scenario. We can also notice the fact the average waiting time is proportional to $T_{max}$.
 
 \subsection{Performance}
We compare the results of the distributed algorithm presented above to other  achieved works \cite{yousef2010intelligent},\cite{faye2012distributed},\cite{faye2012distributed2} on the same isolated-intersection data set of Amiens city in France. Provided the correct value of $T_{max}$, our distributed algorithm achieves the best average waiting time with an average vehicle of $2651$ instead of $2138$ for 3600 steps (3600 s). Moreover, our algorithm is more efficient as it can fast evacuate the emergency vehicles at the intersection while fulfilling the constraints of zero collisions and no green light with an empty queue. In addition to that, this faster evacuation will be  done without significantly increasing the average waiting time of non-emergency vehicles. 
\begin{figure}[!h]
        \centering
         \captionsetup{justification=centering}
        \includegraphics[width=0.48\textwidth]{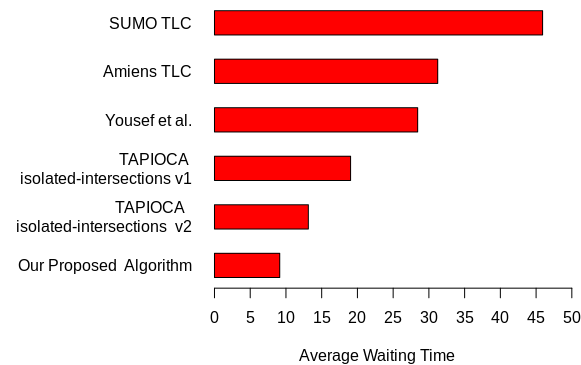}
        \caption{Performance analysis of our result to other result on Amiens isolated dataset \protect\footnotemark. }
        \label{fig :sebastienResult}
\end{figure}
\footnotetext{http://tapioca.sfaye.com/}

%% file: section6-Conclusion.tex
\section{Conclusion and Future Works}
In this work, we presented a vision-based infrastructure of a decentralized approach for the Intelligent Traffic System(ITS).  The main challenges associated with traffic congestion and emergency vehicles were discussed and an adaptive algorithm was presented. \par 
We have modeled an intersection through the conflict matrix, thus giving flexibility in case of a modification of the code of the road, it will be enough to perform an exchange operation on the matrix. The proposed optimized distributed algorithm is based on this matrix and it will provide a set of queues without conflict of collisions and the time needed to empty each queue. We also took into account the balance between the queue with heavy traffic and  low traffic in order to avoid the problem of an empty queue with a  green light.  \par 
We also implemented an algorithm to monitor real-time traffic information using cameras. This low cost and smart vision-based infrastructure approach for gathering traffic data replaces the use of WSN coupled with cameras for video-surveillance which are complex to establish. 

Our future works will attempt to first extend the simulation to multiple intersections data-set while introducing new elements like pedestrian crossing, and secondly port this work on edge devices to perform computation on the edge. 